\documentclass[review]{elsarticle}

\usepackage{lineno,hyperref}
\usepackage{cleveref} % added package!
\usepackage{graphicx, rotating}%, mwe} 
\usepackage{placeins} 
\usepackage{subcaption} % added package!
\usepackage{tablefootnote} % added package!
\usepackage{threeparttable} % added package!
\usepackage{color}  % added package!
\journal{Nuclear Instruments and Methods in Physics Research Section A}

\biboptions{longnamesfirst,semicolon,sort&compress}
\begin{document}
%\graphicspath{{figures/}}

\begin{frontmatter}

%%\title{Scattering neutron background in Multi-Grid detector (worktitle)}
\title{Scattered neutron background in thermal neutron detectors}
%% Modified title baised on Anton's recommendation
%%\title{Investigation of the Scattering Neutron Background of the Multi-Grid Detector Using a Geant4 Model for Thermal Neutron Scattering}
%% Group authors per affiliation:
% \author{Elsevier\fnref{myfootnote}}
% \address{Radarweg 29, Amsterdam}
% \fntext[myfootnote]{Since 1880.}

%% or include affiliations in footnotes:
\author[mymainaddress,mysecondaryaddress,mytertiaryaddress]{E.~Dian\corref{mycorrespondingauthor}}
\cortext[mycorrespondingauthor]{Corresponding author}
\ead{dian.eszter@energia.mta.hu}
\author[mysecondaryaddress]{K.~Kanaki}
%%\author[mytertiaryaddress]{Sz. Czifrus}
\author[myquaternaryaddress]{G.~Ehlers}
\author[mysecondaryaddress,myquinternaryaddress]{R.J.~Hall-Wilton}
\author[mysecondaryaddress]{A.~Khaplanov}
\author[mysecondaryaddress]{T.~Kittelmann}
%\author[mysecondaryaddress]{Sz. T\"{o}r\"{o}k}
\author[mymainaddress,mytertiaryaddress]{P.~Zagyvai}
% \ead[url]{www.elsevier.com}

% \author[mysecondaryaddress]{Global Customer Service\corref{mycorrespondingauthor}}
% \cortext[mycorrespondingauthor]{Corresponding author}
% \ead{support@elsevier.com}

\address[mymainaddress]{Hungarian Academy of Sciences, Centre for Energy Research, 1525 Budapest 114., P.O. Box 49., Hungary}
\address[mysecondaryaddress]{European Spallation Source ESS ERIC, P.O Box 176, SE-221 00 Lund, Sweden}
\address[mytertiaryaddress]{Budapest University of Technology and Economics, Institute of Nuclear Techniques, 1111 Budapest, M\H uegyetem rakpart 9., Hungary}
\address[myquaternaryaddress]{Oak Ridge National Lab, Neutron Technologies Division, Oak Ridge, TN 37831-6475, USA}
\address[myquinternaryaddress]{Mid-Sweden University, SE-851 70 Sundsvall, Sweden}

\begin{abstract}
  Inelastic neutron scattering instruments require very
  low background; therefore the proper shielding for suppressing the
  scattered neutron background, both from elastic and inelastic
  scattering is essential. %DE (KK: write a sentence here that motivates the next one, you don't study shielding here, you identify different sources of background)
  The detailed understanding of the background scattering sources is required for effective suppression.
  The Multi-Grid thermal neutron detector %%, developed at the European Spallation Source
  is an Ar/CO$_2$ gas filled detector with a $^{10}$B$_4$C neutron
  converter coated on aluminium substrates. It is a large-area
  detector design that will equip inelastic neutron spectrometers at the
  European Spallation Source (ESS).  
  To this end a parameterised Geant4 model
  is built for the Multi-Grid detector.
  This is the first time thermal neutron scattering background sources have been modelled in a detailed simulation of detector response.
  The model is validated via comparison with measured data %% from IN6~\cite{khaplanov2014} and CNCS~\cite{khaplanov2017} experiments.
  of prototypes installed on the IN6 instrument at ILL and on the CNCS instrument at SNS.
  The effect of scattering originating in detector components is smaller than effects originating elsewhere. 
  
\end{abstract}


\end{frontmatter}

%%\linenumbers

\section{Introduction}\label{intro}

%%\paragraph{Multi-Grid detector} All Anton's papers, solution for large scale detectors aiming for replacing he3 eg at chopper spectrometers, where the spatial resolution is not crutial. \newline Brief description of general MG geo. eg. frame, blades, coatings, wires!!
Inelastic neutron scattering is a very powerful technique for
exploring atomic and molecular motion, as well as magnetic and crystal
field excitations~\cite{ExpNSc2009}%%\cite{Pharos}
. Time-of-Flight (ToF) %DE(KK: change to ToFveverywhere else in the text from now on)
spectrometers allow a broad phase space
to be measured in a single setting; this is typically achieved with a
large area detector array~\cite{mutka}. In typical state-of-the-art neutron
instruments~\cite{granroth2006, ehlers2011, ehlers2016, mutka,  kajimoto2011, nakajima2011, bewley2011},
this detector array can be
10--50~m$^{2}$. One of the main performance criteria of these spectrometers is typically
defined by the Signal-to-Background Ratio~(SBR), therefore
understanding and enhancing the latter is important for the instrument optimisation. In particular,
scattered neutrons have %%are
a significant contribution to the SBR.
%DE This is
The estimation of the SBR is done currently on a series of prescriptions based on observations of
historical instrument installation. %DE (KK: I don't get the last sentence. Do you mean how the sbr is derived so far?)

%DE \paragraph{The Multi-Grid detector}% is a large-area detector designed for chopper spectrometry with cold and thermal neutrons~\cite{andersen_2012}~\cite{A.Khaplanov2012}.}
%DE (KK: you don't need a paragraph title here, just let the text flow, you break the flow of the intro)

As a consequence of the recent restructuring of the $^{3}$He market~\cite{shea}, a need for cost effective $^{3}$He-replacing detector solutions is raised~\cite{zeitelhack2012}, especially for %% in the case of the
inelastic neutron scattering instruments, where large area detectors with high %%Signal-to-Background Ratio~(SBR)
SBR are required. A potent new solution for this type of instruments
is the Multi-Grid detector~\cite{andersen_2012, A.Khaplanov2012},
which will be used for the three Time-of-Flight chopper spectrometers 
at %dE(KK: I wrote ESS in the abstract, use ESS from now on) the European Spallation Source
ESS~\cite{ess, deen2015, TREXprop, CSPECprop}.
%DE ~\textcolor{red}{(I didn't find paper reference for T-REX and C-SPEC, only the proposals)}.
  The Multi-Grid design was invented
at the Institut Laue-Langevin (ILL)~\cite{mg_patent_ill, ill},
and the detector now is jointly developed by the ILL and
the ESS within the CRISP~\cite{crisp}
%DE ~\textcolor{red}{(check for valid reference...)}
and BrightnESS~\cite{brightness} projects.

The Multi-Grid detector is an Ar/CO$_{2}$-filled proportional chamber
with a solid boron-carbide ($\rm ^{10}B_{4}C$) neutron converter,
enriched in $^{10}$B~\cite{hoglund2012, c.hoglund2015b, LiU2016b}. %DE \textcolor{red}{(Which 3 LiU papers?)}.
%%REVISED
{The basic unit of the Multi-Grid detector is the grid, an aluminium frame; thin aluminium lamellas, coated on their both sides with boron-carbide, the so called blades are placed in this frame, parallel with each other and the entrance window of the grid, dividing the grid into cells. In the detector the grids are structured into columns, and this way the cells one above the other form tubes, and the signals are readout both from the frames and the anode wires that go through the whole length of the column in the centre of the cells. The planned detector modules and the prototypes are built of these columns. %%
A series of small
size prototypes and large scale demonstrators are already built and
tested at different sources and instruments~\cite{khaplanov2014,
  khaplanov2017}, and the development of the detector has already
entered the up-scaling phase. As Multi-Grid is a large area detector, full
scale design is limited by cost considerations. However, detailed
Monte Carlo modelling can help tackle the limitations and provide
guidelines for the up-scaling design, which is particularly important
for detectors that have to provide excellent $\rm SBR \sim \mathcal{O}(10^{5})$.
%DE(KK: get this number from Pascale or Anton, we have to mention it once)

The two-fold aim of the current study is to introduce a detailed
Geant4 model of the Multi-Grid detector including validation against
datasets of experiments~\cite{khaplanov2014, khaplanov2017} performed
on existing %DE built %DE(KK: existing?)
demonstrators, as well as to identify the
various components of the scattered neutron background, induced by
cold and thermal neutrons %DE in the inner and the outer environment of
internally and externally to the Multi-Grid detector. %DE (KK: internally and externally to the detector??).

%%In this paper in Section~2 a detalied and flexible Geant4 model of the Multi-Grid detector is presented.
%%The model was validated against measured ToF and energy transfer data of the demonstrator detectors tested at the ILL and the CNCS. The built mondel and the compariosn of the measured and simulated results are showed in Section~3. 

The Geant4 model of the Multi-Grid detector is presented in
Section~\ref{appmet}. In Section~\ref{illmod}, %%DE ,illsim,cncsmod,cncsRnToF,cncsEtrf}
the model validation against the measured ToF %DE Time-of-Flight,
flight distance and energy transfer data from the IN6~(Cold neutron time-focusing time-of-flight spectrometer IN6-Sharp), and in~\Cref{cncsRnToF,cncsEtrf} the CNCS~(Cold Neutron Chopper Spectrometer)  
demonstrator tests are shown. %DE (KK: spell the instrument names out first time you use them, earlier I think).
As part of the reproduction 
of the CNCS demonstrator measured data, a study of the
individual %%GEcomponents %DE (KK: don't get eliminated, what do you mean here? study of background reduction?)
%%GEof scattering
contributions to the scattered neutron background is also discussed. In
Section~\ref{cncswin} results regarding the neutron scattering on the
aluminium components of the detector vessel of the CNCS detector are
described. Finally, in Section~\ref{conc} the obtained results are
concluded from the aspects of validation, and the further utilisation
of the built model for detailed background analysis and for the
optimisation of the detector vessel design is also shown.
\section{Geant4 model of Multi-Grid detector}\label{appmet}

%% \textcolor{blue}{(KK: what is important for this section?
%% 1. Geometry, you have the table
%% 2. MAterials, you mention NXSG4, you can say the rest are from the G4
%% material database
%% 3. Physics lists: what I say below
%% 4. generators, you have a dedicated paragraph)}

A general, parameterised Geant4~\cite{agostinelli, allison2006,
  allison2016} model of the Multi-Grid detector has been
developed within the ESS Detector Group Simulation
Framework~\cite{kittelmann2013b}, with the usage of the
NXSG4~\cite{kittelmann2015b} extension library.
%DE The latter offers a detailed and realistic treatment of low energy neutrons, so incoherent and
%DE coherent scattering for poly-crystalline materials (i.e.\,the aluminium detector frames)
%DE are taken into account.
%DE (KK: the last sentence is wrong, remove it.)
%%The NXSG4 is applied in the definition of the following materials in the models: the aluminium, used for the detector frame, the B$_{4}$C coating and the polyethylene, which was an estimation of epoxy glue in some of the shielding materials. For all other components, standard Geant4 materials are used.
The latter enables the crystalline structure of aluminium, used in the
detector frame. For all other components, standard Geant4 materials
are used. The physics list is the standard QGSP\_BIC\_HP, except
when polyethylene is included in the materials, in which case a
customised physics list is preferred instead~\cite{kittelmann2017}, due to the relevance of thermal scattering on the high hydrogen-content of the polyethylene.
% was an estimation of epoxy glue in some of the shielding materials. Due to the relevance of thermal scattering on the high hydrogene-content of the polyethylene, an internal, customised list~\cite{kittelmann2017}, the QGSP\_BIC\_HP\_TS is applied.

%DE
%% \textcolor{blue}{(KK: the entire next paragraph except where you introduce the table,
%% last sentence, is
%% boring :-) nobody cares how you did it. You will never see this
%% section even in G4 papers. We need to discuss your physics list. You
%% should cite the G4 one and say that we use a custom one for the PE scattering)}
%%!!In the simulation the custom QGSP\_BIC\_HP physics list is used for the polyethylene, and an internal, customised list~\cite{kittelmann2017} is applied for the rest of the materials.
From the flexible, full-scale model, the realistic models of two demonstrators that were tested at the IN6~\cite{khaplanov2014} at ILL and and at the CNCS~\cite{khaplanov2017} at SNS were also prepared.
\begin{figure}[ht!]
  \centering
  \begin{subfigure}[b]{0.32\textwidth}
    \includegraphics[width=\textwidth]{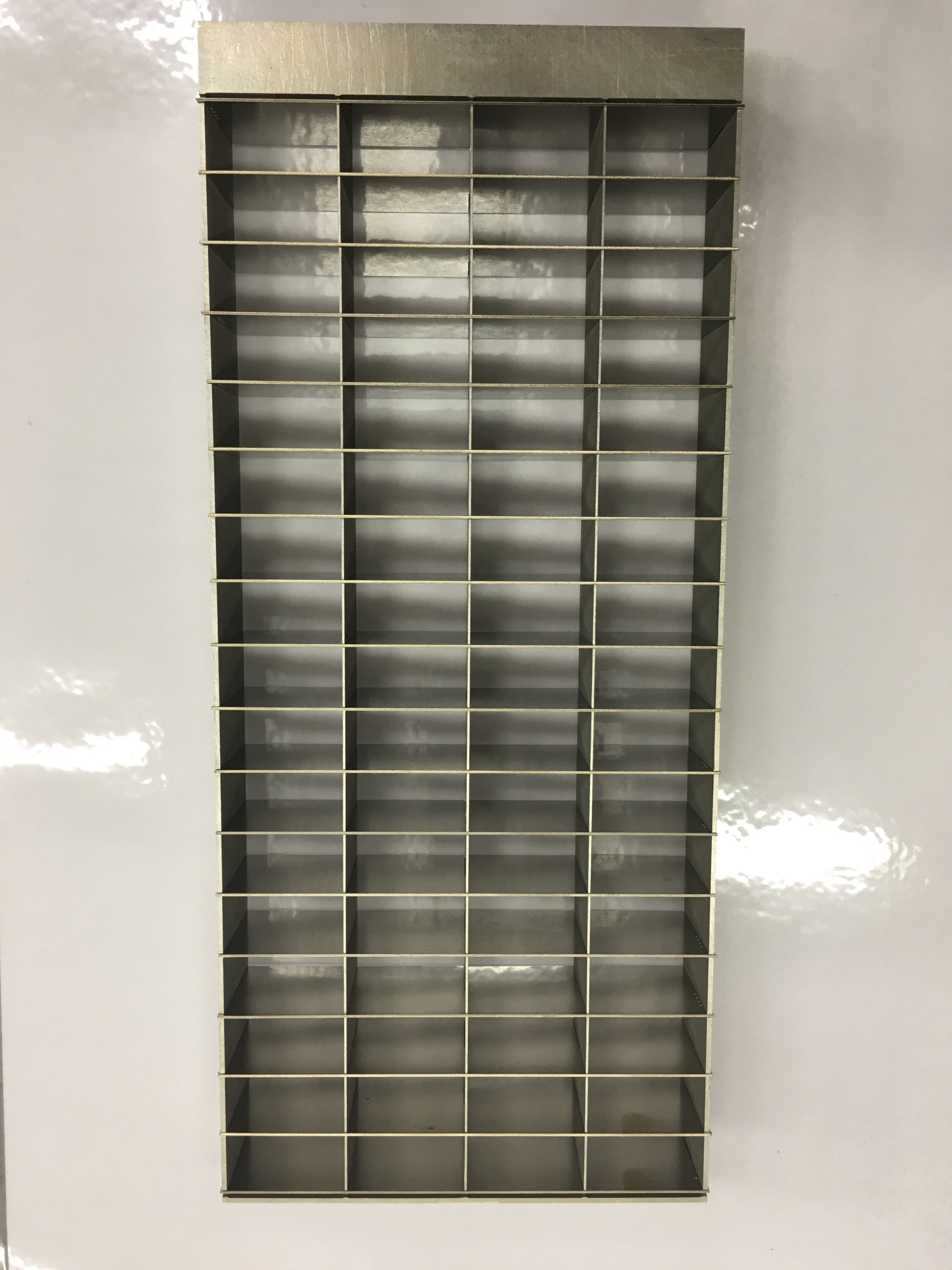}
    \caption{\label{gridreal}}
  \end{subfigure}
  \begin{subfigure}[b]{0.2\textwidth}
    \includegraphics[width=\textwidth]{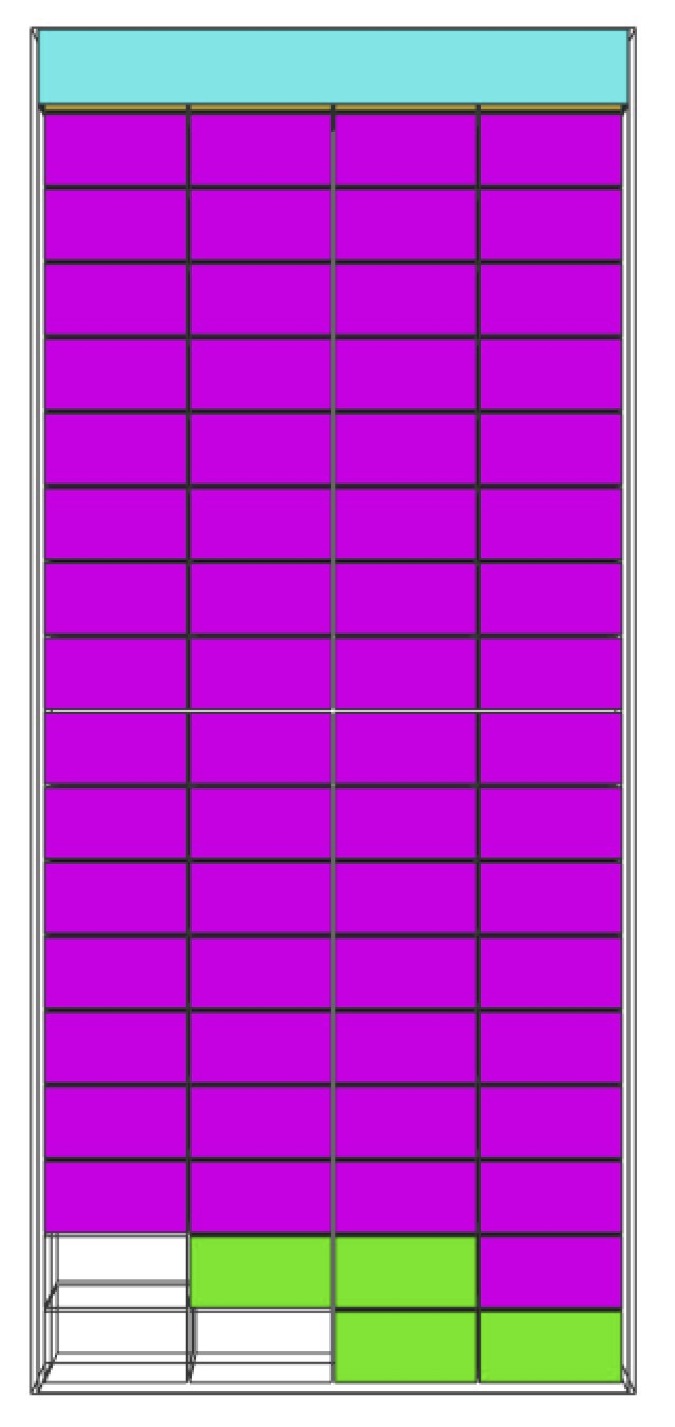}
    \caption{\label{gridsim}}
  \end{subfigure}

  \caption{Real grid~(\ref{gridreal}) and grid geometry implemented in Geant4~(\ref{gridsim}).  \label{gridgeom} }  
\end{figure} 
To reach a very flexible geometry, a few simplifications were done. The basic unit of the model is the so called cell, a $\rm 2 \times 2 \times 1$~cm$^{3}$ counting gas volume of the detector, delimited by B$_{4}$C-coated aluminium blades. Therefore everything has to be symmetrical at the cell level, like the blade thicknesses, the cell volumes and most importantly the coating thicknesses. This estimation is applied in the basic model and the IN6, but in the CNCS demonstrator a series of different coating thicknesses are used ($\rm 13 \times 0.5~\mu m + 14 \times 1.0~\mu m + 6 \times 1.5~\mu m + 1 \times 1.0~\mu m$). So for the latter model, the coating thicknesses are hard-coded to fit the real prototype, and so %%like this
the grid became the basic unit for this detector. The anode wires and the electronics of the detector are excluded from %%neglected within
the models, %, and the end front of the grid was defined as one aluminium box, not giving spaceholder for the electrical conjunctions
as it is shown in Figure~\ref{gridgeom}. The major parameters of the prepared models are shown in Table~\ref{tab:geomparam}. 

%%The materials of the specific volumes also were parametrically defined. For the aluminium frame and the B$_{4}$C specific material compositions were used, being pre-defined at the ESS Coding Framework. For this materials the neutron scattering properties given by the afore mentioned NXSG4 and NCrystal tool, therefore their crystal structure considered in the simulation. Same type of polyethylene was used for the end shielding of the detectors, where gadolinium was applied mixed with acrylic painting, that was estimated as 50~\% Gd$^{2}$O$^{3}$ and 50~\% polyethylene. All other materials were built from standard G4 materials ~[god ref needed here!]. The major parameters of the prepared models are shown in Table~\ref{tab:geomparam}. 

\begin{table}[htbp]
  \centering
  \caption{Major default geometrical parameters of Multi-Grid detector models.}
  \label{tab:geomparam}
  \resizebox{\textwidth}{!}{
    \begin{threeparttable}
      \begin{tabular}{llccc}
        \hline
        \multicolumn{2}{c}{Parameter}                    &  \multicolumn{3}{c}{Default value}                         \\
                                      &                  & Basic model    & IN6 model           &      CNCS model     \\
        \hline
        Number of cells               & width~(x)        &   4$^{\phantom{1}}$            &  4                  & 4                   \\
                                      & depth~(z)        &  17$^{\phantom{1}}$            & 17                  & 17                  \\
        Number of grids in columns    &                  & 127$^{\phantom{1}}$            & 16                  & 48                  \\
        Number of columns             &                  & 125$^{1}$      &  6                  &  2                   \\
        Cell size~                    & width~(x)        &   2.2~cm       &  2.2~cm             &  2.2~cm             \\
                                      & height~(y)       &   2.26~cm      &  2.26~cm            &  2.25~cm            \\
                                      & depth~(z)        &   1.1~cm       &  1.1~cm             &  1.1~cm             \\
        Coating thickness             &                  &   1.0~$\mu$m   &  1.0~$\mu$m         &  0.5-1.5~$\mu$m     \\
        Entrance window thickness     &                  &   1.0~mm       &  1.0~mm             &  2.0~mm             \\
        Frame end thickness           &                  &  11.6~mm       & 11.6~mm             & 12.5~mm             \\
        Frame side thickness          &                  &   1.0~mm       &  1.0~mm             &  1.0~mm             \\
        Blade thickness               & orthogonal~(z)   &   0.6~mm       &  0.6~mm             &  0.5~mm             \\
                                      & parallel~(x)     &   0.5~mm       &  0.5~mm             &  0.5~mm             \\                       
        End shielding thickness       &                  &   1.0~mm       & 10$^{-7}$~mm$^{2}$         &  1~mm               \\
        Side shielding thickness      &                  &   1.0~mm       &  0~mm               &  0~mm               \\
        Grid gap thickness            &                  &   1.0~mm       &  1.0~mm             &  1.0~mm             \\
        Sample-detector front face distance      &                  &   4~m          &  2.48~m        &  3.33~m       \\ %DE 3.33522~m          \\
        \hline
        %DE Physics list                  &                  & ESS\_QGSP\_BIC\_HP\_TS  & ESS\_QGSP\_BIC\_HP\_TS  & ESS\_QGSP\_BIC\_HP\_TS \\
        %DE Physics list                  &                  & \multicolumn{3}{c}{ESS\_QGSP\_BIC\_HP\_TS}                 \\
        Physics list                  &                  & \multicolumn{3}{c}{QGSP\_BIC\_HP}                 \\
        \hline
        %DE Frame material                &                  & Al$^{3}$      & Al$^{3}$            &  Al$^{3}$          \\
        Frame material                &                  & Al             & Al                  &  Al                  \\
        Counting gas                  &                  & Ar/CO$_{2}$    & Ar/CO$_{2}$          &  Ar/CO$_{2}$        \\
                                      &                  & 80/20          & 90/10               &  80/20              \\
        Coating                       &                  & $^{10}$B$_{4}$C &  $^{10}$B$_{4}$C     & $^{10}$B$_{4}$C      \\
                                      &                  & 97~\% enriched & 97~\% enriched      & 97~\% enriched       \\
        
        End shielding                 &                  &  PE+Gd$_{2}$O$_{3}$         &  -                  &  PE+Gd$_{2}$O$_{3}$              \\
        \hline
        Modules                       &                  & no             & no                  & yes                 \\
        Vessel                        &                  & -              & -                   & yes                 \\
        \hline
      \end{tabular}
      \begin{tablenotes}\footnotesize
      \item $^{1}$Number of columns defined to build a typical 180$^{\circ}$ detector arc.
      \item $^{2}$End shielding is implemented as a volume of
        PE+Gd$_{2}$O$_{3}$, therefore 0~mm thickness is not allowed by
        the code. Lack of shielding was obtained with the minimum
        applicable thickness. %DE \textcolor{blue}{(KK: again nobody cares....)}
        %DE
      %% \item $^{3}$Crystalline aluminium enabled with NXSG4.  \textcolor{blue}{(KK:
      %%   already said it. ACtually none of these footnotes are really
      %%   important, perhaps only 1)}
      \end{tablenotes}
  \end{threeparttable}}
\end{table}
%% \footnotesize{$^{*}$Number of columns defined to buld a typical 180 detector arc}
%% \footnotesize{$^{**}$Crystalline aluminium built with NXSG4}

%For each model, the source of the neutrones was a the sample position, neglecting the scattering on the sample itself. (Maybe reasoning?) The
%Within the simulations, the neutrons were run from sample position, and all secondary particles were also considered. The sample was defined at the center og the geometry, while z direction was chosen as the beam direction, leading to x as horisontak and y as vertical coordinates. Beside the basic particle guns, like pencil beam, isotropic point or cylindrical sources, some more realistic cases were as well applied. Although the scattering on the sample was itself was neglected, some sample and instrument effects were added via the source definition, like the volume of the sample, the initial energy or Time-of-Flight (ToF) distribution of the neutrons. 

%A few simplifications were applied for the simulated neutron sources as well.
The simulated primary neutrons are generated at the sample position. The sample is placed at the
centre of the geometry, with the $z$ direction chosen as the beam
direction, leading to $x$ as horizontal and to $y$ as vertical
coordinates. The sample-to-detector distance is defined as the
shortest distance from the sample position to the entrance window of the detector: grid window
or vessel window, in case the latter is enabled. Basic particle guns, like a pencil beam,
%DE isotropic
$\rm 4 \pi$ and cylindrical sources are used, as well as targeted beams
to irradiate only the detector surface. Although the physics of the
samples themselves is not implemented in the simulations, the above
listed particle guns are defined both as point and volume sources
(1~x~1~x~1~cm$^{3}$ cube or cylinder with 1~cm diameter). Some
instrument effects are introduced via the source definition, like the
energy distribution %DE or the initial Time-of-Flight distribution
of the incident primary neutrons. 

%DE The models are validated against and the scattered neutron study were performed via the same quantities that can be measured with real demonstrators: detection coordinates, ToF, and the calculated energy distribution and energy transfer.

In chopper spectroscopy the data of interest are the momentum- and energy transfer of the
scattered neutrons. These are derived from the primary measured
quantities: the detection coordinates (giving the flight distance) and the
ToF. The flight distance is defined as the distance from the sample
position to the detection coordinates. The
simulated detection coordinates are %%similarly
reduced to the centre of
the cell in which the neutron is detected, despite the higher
resolution of the simulation. ToF is simulated from sample position.
The detector model is validated against these raw measured quantities of the IN6 demonstrator, and a detailed study of scattered neutron background is also performed regarding the energy transfer in the CNCS demonstrator.

%%The great benefit of the Monte Carlo simulations, that all data of the particle tracks can be recorded. Therefore, on the one hand, all data can be simulated that are usual during the measurements, like ToF, detection coordinates, energy deposition, measured energy transfer, etc., but on the other hand, all of these can be followed in particle level, so the sources of the different effects, like neutron scattering, neutron escape can be determined. In the current study, neutron ToF, conversion coordinates, and detection coordinates were simulated, and energy transfer was derived from simulated quantities. (\textbf{Maybe extend later})

%textbf{I decided to change the structure, not describeing searately the the experiment and the prototype-model, O think it's simplier this way after a detailed presentation of the basic bodel and the parameters.}

%DE \section{Results \& Discussion, KK: again why not call it IN6 results?}\label{RnD}  
% \section{IN6 results}\label{RnD}

%%\subsection{The IN6 experiment} %The model of the Multi-Grid prototype used at the IN6 experiment is a detector of 6 columns, with 16 grids in each. For this detector
%DE
%% \subsection{Model of demonstrator test on IN6 instrument at ILL, \textcolor{blue}{KK::
%%   find a concise title}} \label{illmod}
\section{Model of demonstrator test on the IN6 instrument at ILL} \label{illmod}
%DE
%% At the IN6 experiment, a 6-column Multi-Grid prototype was
%% tested~\cite{khaplanov2014}. The detector is built up from 6~x~16
%% grids, 4~x~17 cells in each grid (see Figure~\ref{gin6_real} and
%% Table~\ref{tab:geomparam}), with no shielding at the rear end of the
%% grids. The detector is filled with Ar/CO$_{2}$ (90/10
%% by volume) at %% $\rm p = 1~bar$
%% nominal room temperature and pressure. The distance from the sample
%% position to the front surface of the grids is 248~cm. (KK: again, is
%% all this necessary when the table and the figures tell you all you need
%% to know?)
At the IN6 experiment the demonstrator~(Figure~\ref{gin6_real}) 
%DE The detector
is tested with neutron beams of 4.1, 4.6 and 5.1~\AA~(i.e. 4.87, 3.87 and 3.15~meV, respectively),
irradiating the entire entrance surface.
%DE  The same geometry is built for simulation, as it is shown in Figure~\ref{gin6_g4}, in order to reproduce the published ToF spectra, and ToF spectra as the function of the depth of detection, to validate built geometry.
The same geometry is implemented in the simulation~(Figure~\ref{gin6_g4}) and validated against the measured and published ToF spectra.
%DE (KK:make this sentence concise).
Due to the lack of data on the measurement
setup (e.g., exact chopper %DElocation
settings and timing references), the measured and simulated ToF spectra are compared
 either in a relative time scale, or all of them are scaled to the time
scale of the simulation, in which the neutrons and their respective
ToF are generated at the sample position.  
\FloatBarrier
\begin{figure}[ht!]
  \centering

  \begin{subfigure}[b]{0.49\textwidth}
    \includegraphics[width=\textwidth]{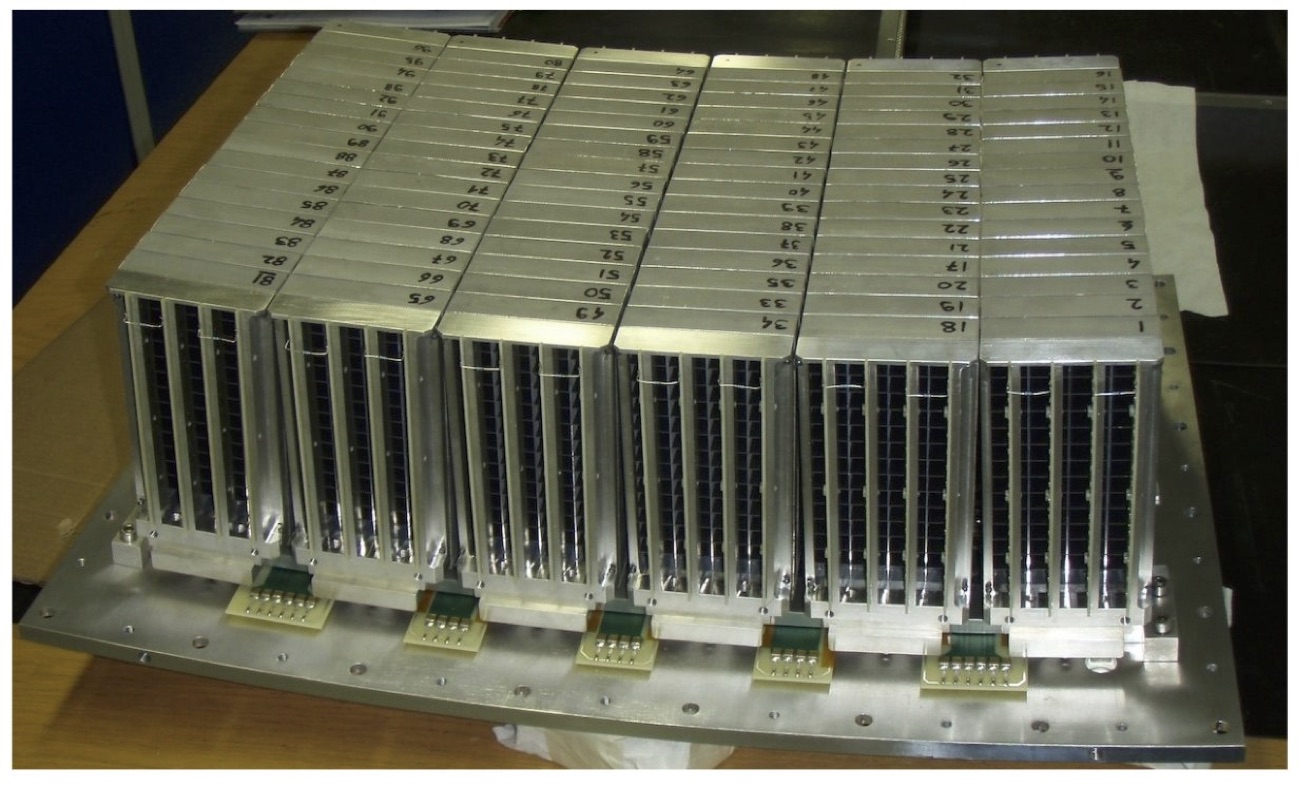}
    \caption{\label{gin6_real}}
  \end{subfigure}
  \begin{subfigure}[b]{0.49\textwidth}
    \includegraphics[width=\textwidth]{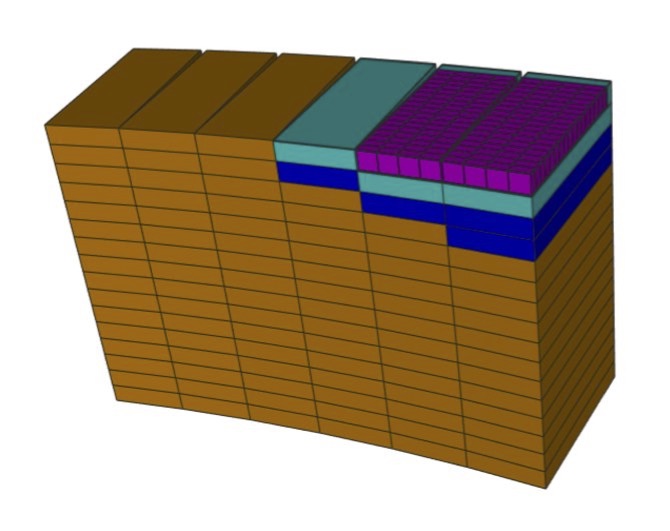}
    \caption{\label{gin6_g4}}
  \end{subfigure}

  \caption{As built IN6 prototype~(\ref{gin6_real}) and its Geant4 model~(\ref{gin6_g4}).  \label{gin6}}
\end{figure}

The detector geometry is irradiated with pencil and targeted beams, in
order to illuminate the entrance surface (see Figure~\ref{beamin6}),
both with sharply mono-energetic and Gauss-smeared initial neutron energy
distributions of 4.1, 4.6 and 5.1~\AA. For preparing the demonstrative
study on the 2-dimensional distributions of the ToF spectra
as the function of the depth of detection, a minor simplification was
performed: for this demonstration only 1 column of the detector model
was used, since in this case z-coordinate one-to-one corresponds to the detection depth in detector, leading to an easy readout.
\FloatBarrier
\begin{figure}[ht!]
  \centering
  \includegraphics[width=\textwidth]{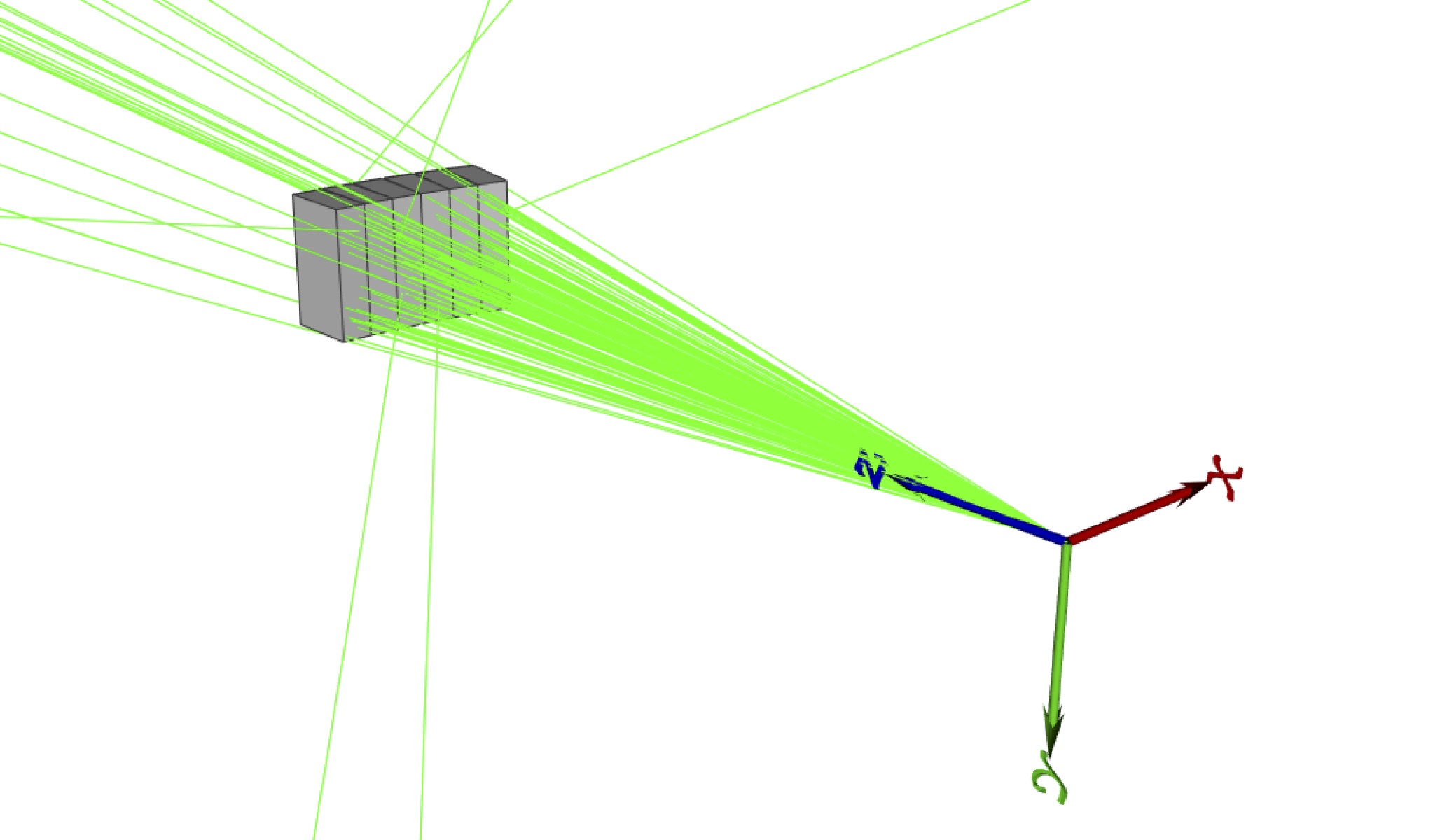}
  \caption{Geometry view of the IN6 Geant4 detector model irradiated
    with targeted beam.
    \label{beamin6} }
\end{figure}

%%\subsection{The IN6 results}
\subsection{Simulation results for IN6 Demonstrator detector}\label{illsim}

For the IN6 experiment, ToF spectra and 2D detection depth
dependent ToF spectra are simulated %are performed
and compared to the published %%measured ones
measurements at 4.1, 4.6 and 5.1~\AA \ wavelengths.
In Figure~\ref{in6_2d} the comparison of the measured 
%% REVIRED
(\Cref{2D_m_4.1,2D_m_4.6,2D_m_5.1}) and the simulated  ToF-spectra as a function of the depth of detection is presented with
mono-energetic (\Cref{2D_s_4.1_mono,2D_s_4.6_mono,2D_s_5.1_mono}) and Gaussian (\Cref{2D_s_4.1_gauss,2D_s_4.6_gauss,2D_s_5.1_gauss}) incident neutron energy distributions. At all wavelengths the main path of
the %DE primarily
incident detected neutrons clearly appears as a skew line both in
the measured and the simulated distributions. The angle of the path is related to the neutron's velocity. 

Beside the main path, at 4.1 and 4.6~\AA \ wavelengths that are below the
aluminium Bragg edge~\cite{ref4Bragg, AlCrystal}, the traces of the
detected scattered neutrons appear as well. %%too.
On the one hand, in the near
surface region a triangle-shaped shadow appears beside the main
neutron path, produced by the neutrons detected after scattering on
the intermediate aluminium blades. On the other hand, a short,
opposite direction skew line appears for these two wavelengths, both in
the measured and simulated distributions, starting from the unshielded
rear end of the detector, caused by a significant fraction of
scattered neutrons coming from the detector end blade. Both effects
are caused by the Bragg-scattering on aluminium and emphasise the need for targeted shielding in the detector.

%DE
%% (KK: careful with your grammar, rewrite it to pass to the next
%% paragraph, create flow.) As all these Time-of-Flight characteristics and scattering phenomena are successfully reproduced. This serves as a qualitative validation of the developed Geant4 model.
With the reproduction of these ToF characteristics and scattering phenomena, the developed Geant4 model is qualitatively validated. For a quantitative validation, 1D ToF histograms are also simulated.
%%
%% %2D plots
%% \begin{figure}[ht!]
%%   \centering
%%   \begin{subfigure}[b]{1\textwidth}
%%     \includegraphics[width=\textwidth]{mg_plots/measured_4p1.png}
%%     \caption{\label{2D_m_4.1}}
%%   \end{subfigure}
 
%%   \begin{subfigure}[b]{1\textwidth}
%%     \includegraphics[width=\textwidth]{mg_plots/simulated_4p1.png}
%%     \caption{\label{2D_s_4.1}}
%%   \end{subfigure} 
%%   \caption{ToF spectra as the function of the depth of detection. Results of measuement at the IN6 experiement (\ref{2D_m_4.1}, measured data taken from~[ref Anton's paper]) and Geant4 simulation at simulation (\ref{2D_s_4.1}) at 4.1~\AA. \label{in6_2d}}
%% \end{figure}
%%\FloatBarrier %new
%%2D plots
\begin{sidewaysfigure}[ht!]
  \centering
  
  \begin{subfigure}[b]{0.32\textwidth}
    \includegraphics[width=\textwidth]{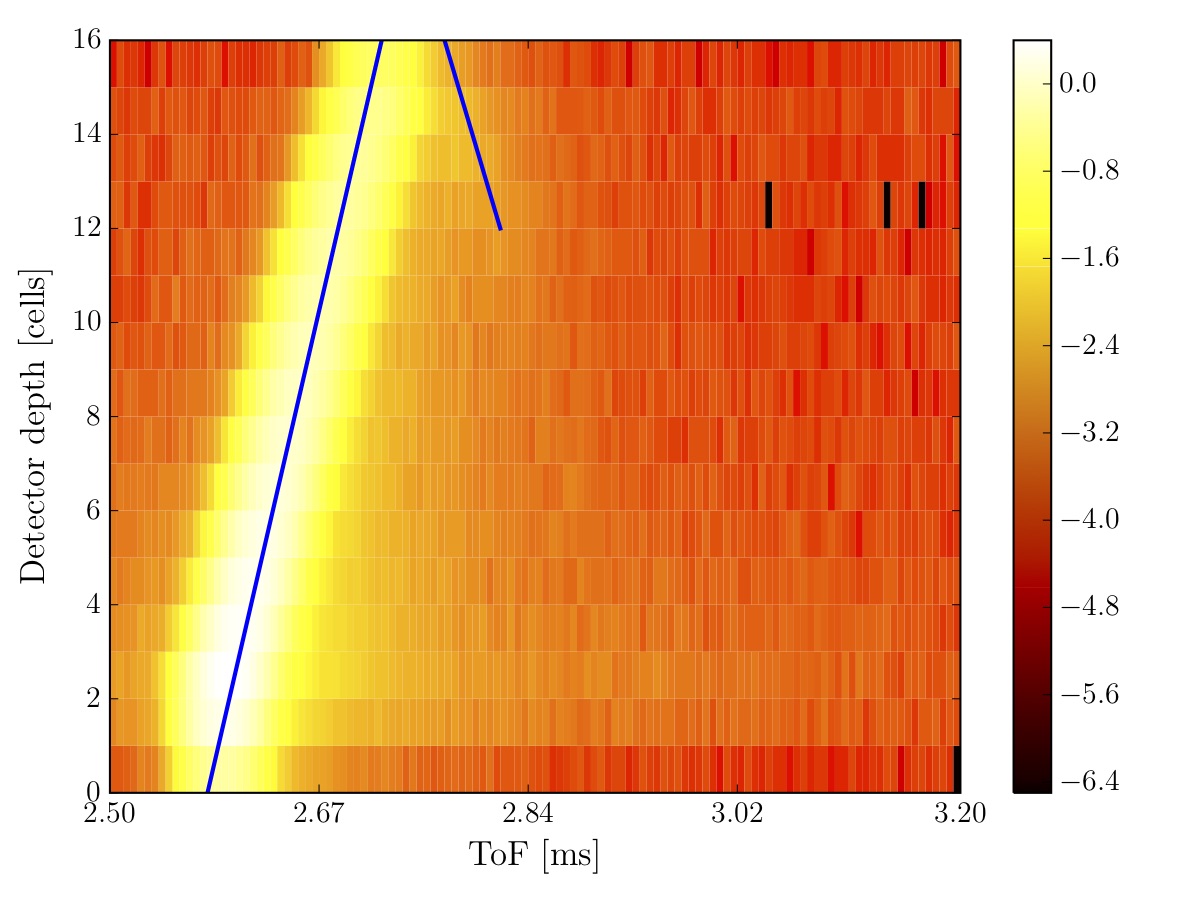}
    \caption{Measured ToF at 4.1~\AA. \label{2D_m_4.1}}
  \end{subfigure}
  \begin{subfigure}[b]{0.32\textwidth}
    \includegraphics[width=\textwidth]{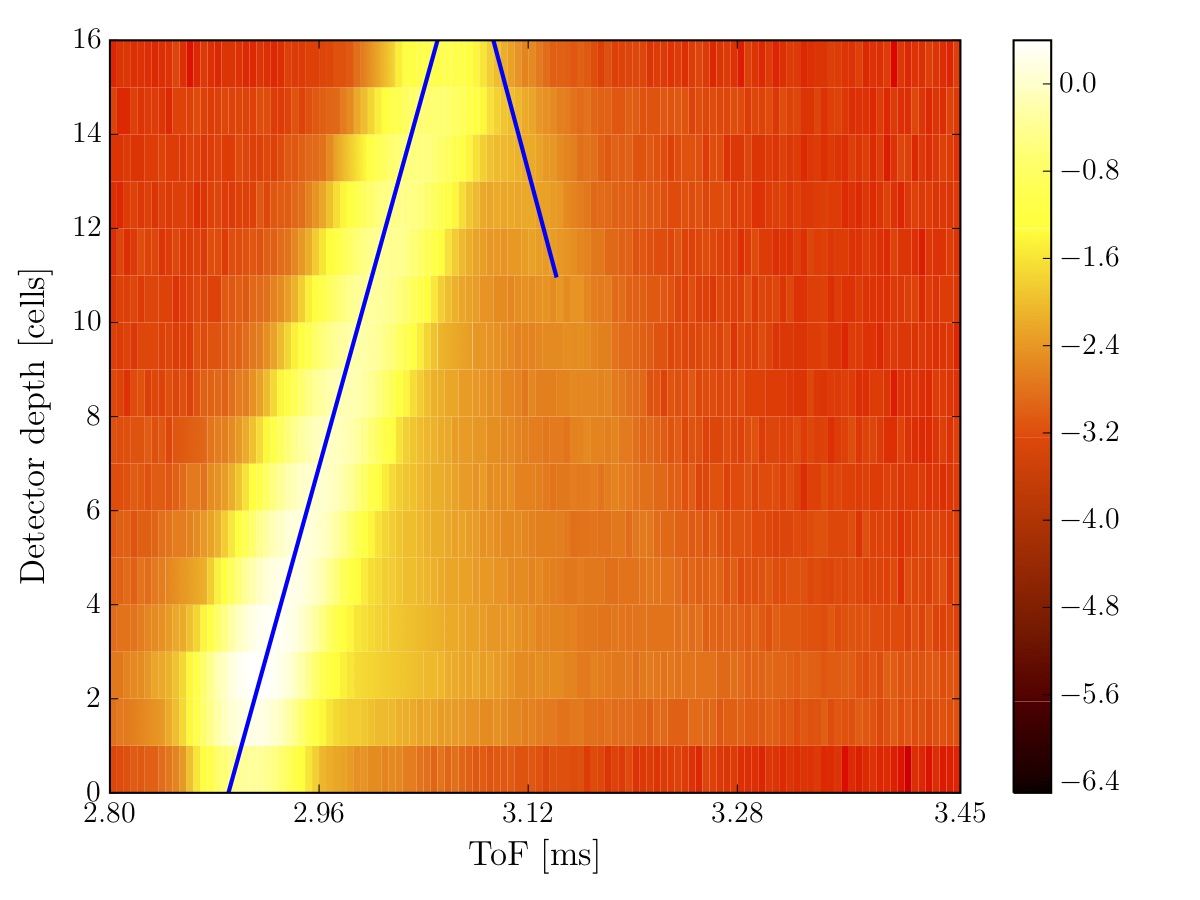}
    \caption{Measured ToF at 4.6~\AA. \label{2D_m_4.6}}
  \end{subfigure}
  \begin{subfigure}[b]{0.32\textwidth}
    \includegraphics[width=\textwidth]{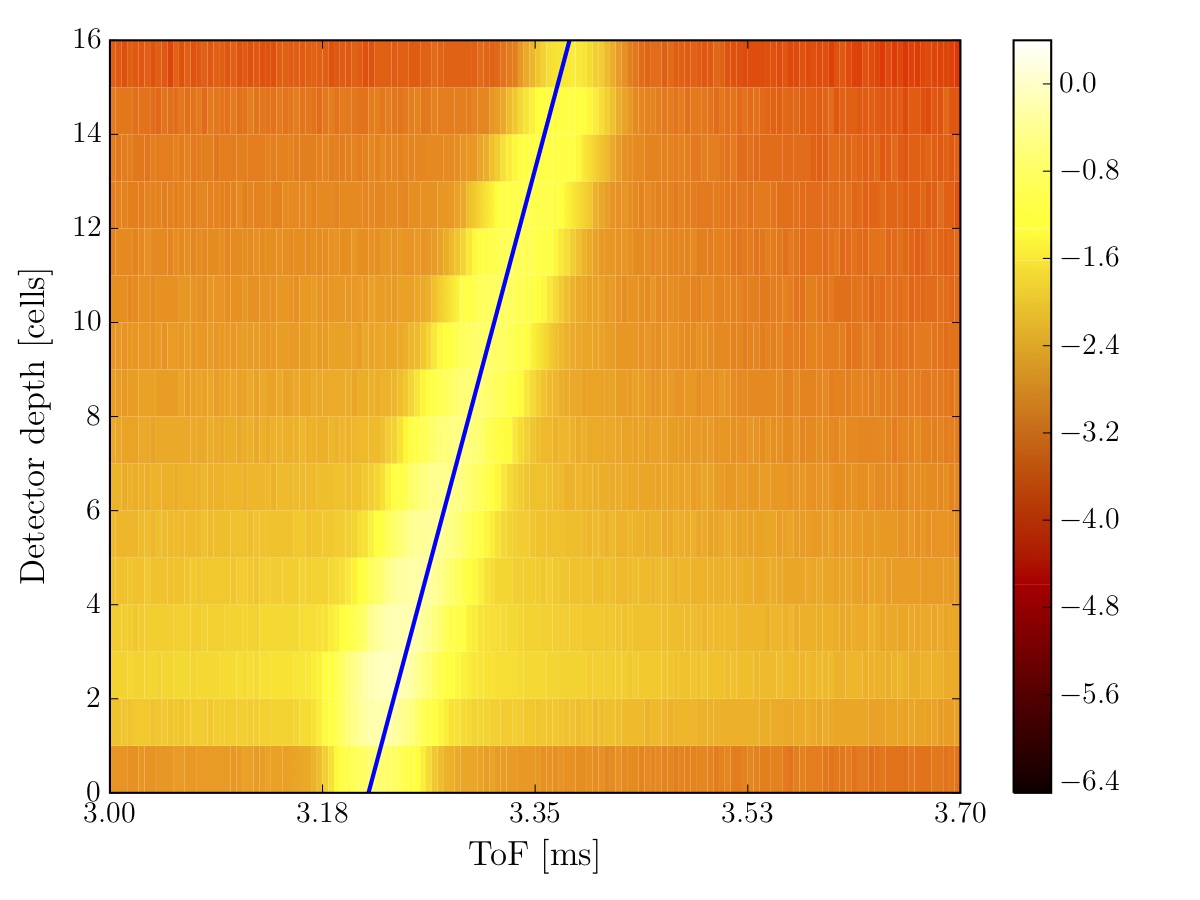}
    \caption{Measured ToF at 5.1~\AA. \label{2D_m_5.1}}
  \end{subfigure}
  
  \begin{subfigure}[b]{0.32\textwidth}
    \includegraphics[width=\textwidth]{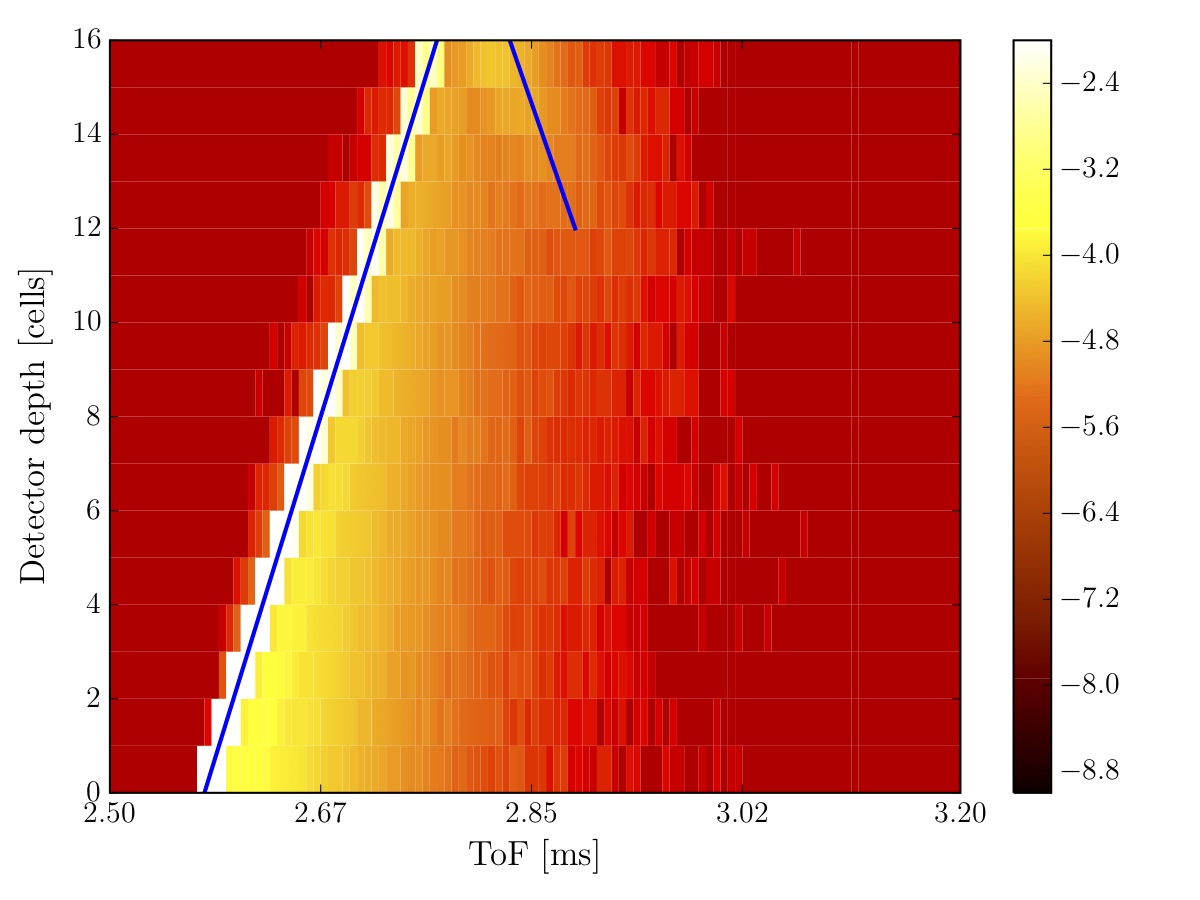}
    \caption{Simulated ToF at 4.1~\AA. \\ E$_{ini}$: mono-energetic \label{2D_s_4.1_mono}}
  \end{subfigure}
  \begin{subfigure}[b]{0.32\textwidth}
    \includegraphics[width=\textwidth]{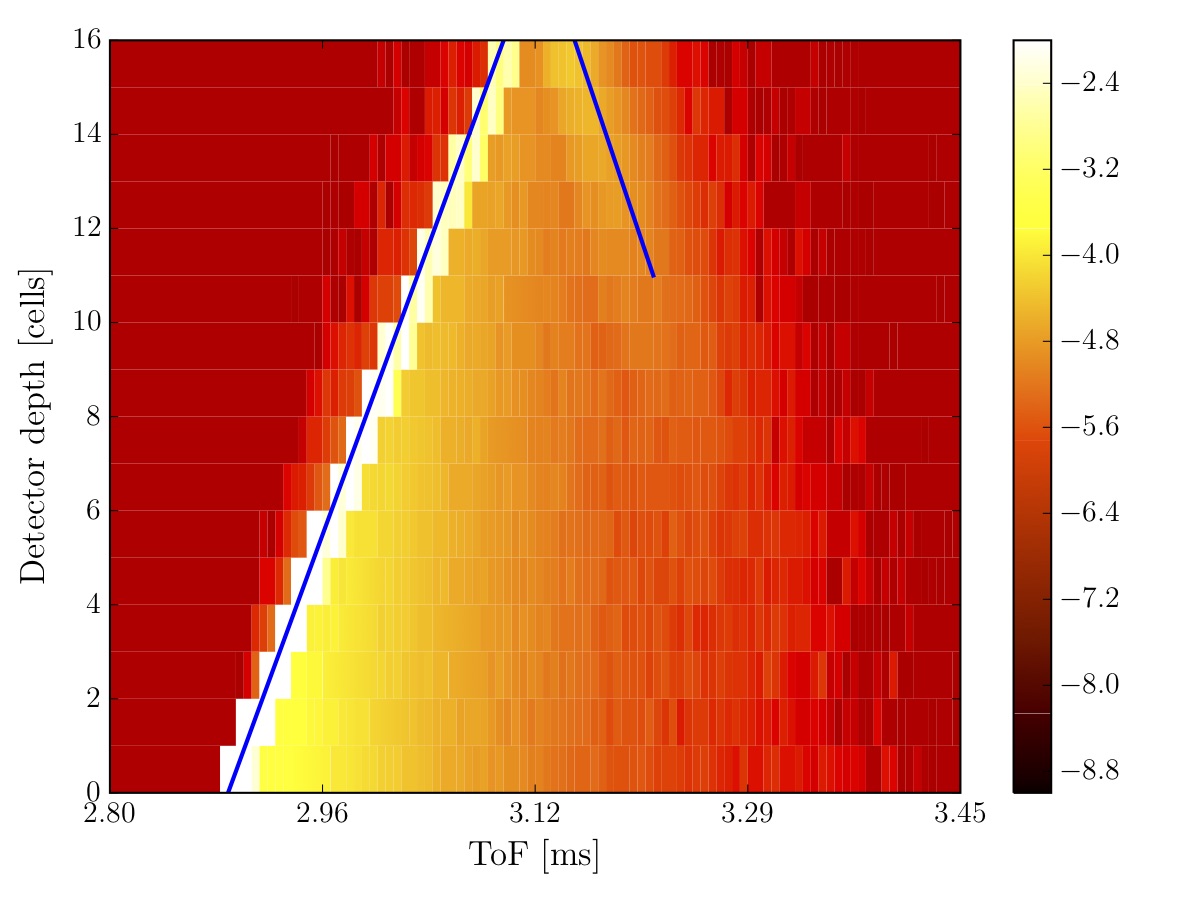}
    \caption{Simulated ToF at 4.6~\AA. \\ E$_{ini}$: mono-energetic \label{2D_s_4.6_mono}}
  \end{subfigure}
  \begin{subfigure}[b]{0.32\textwidth}
    \includegraphics[width=\textwidth]{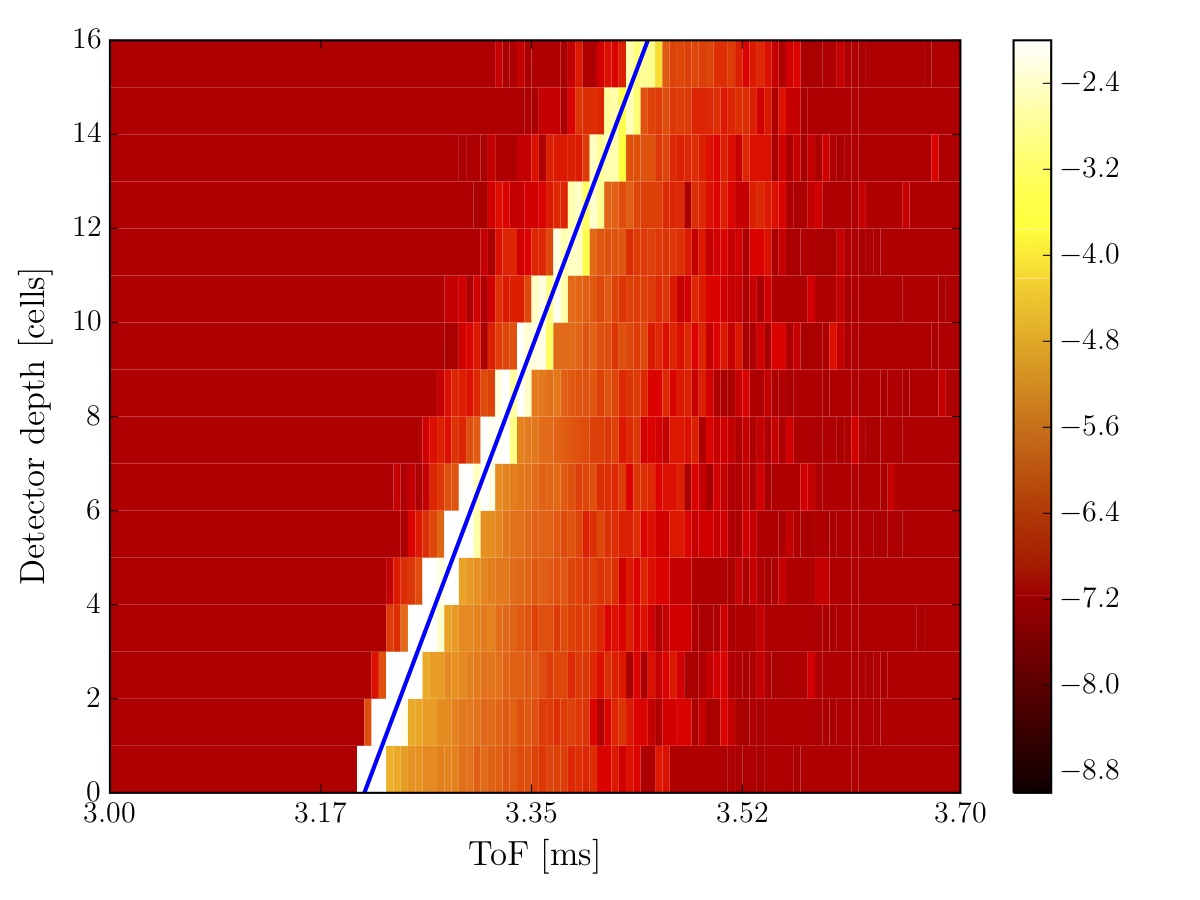}
    \caption{Simulated ToF at 5.1~\AA. \\ E$_{ini}$: mono-energetic \label{2D_s_5.1_mono}}
  \end{subfigure}
    
  \begin{subfigure}[b]{0.32\textwidth}
    \includegraphics[width=\textwidth]{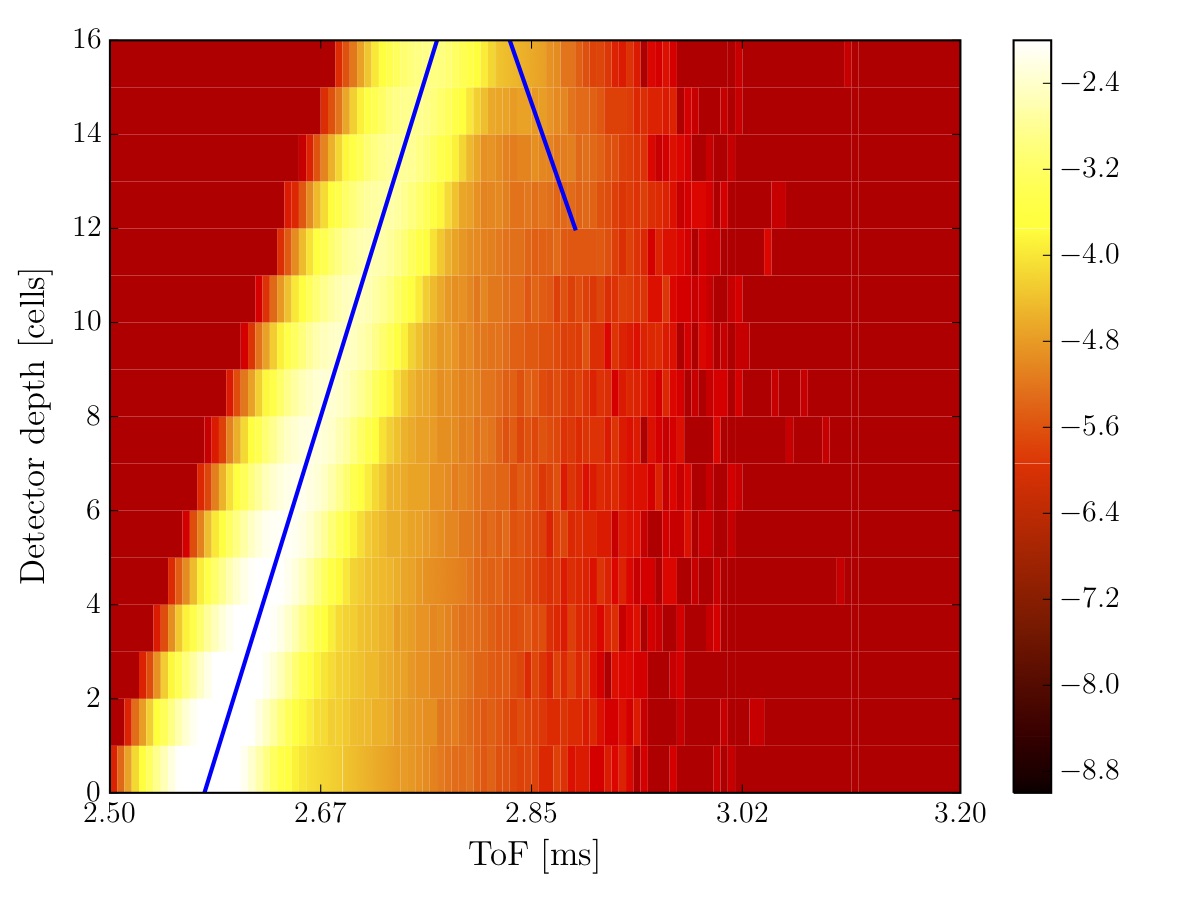}
    \caption{Simulated ToF at 4.1~\AA. \\ E$_{ini}$: Gaussian \label{2D_s_4.1_gauss}}
  \end{subfigure}
  \begin{subfigure}[b]{0.32\textwidth}
    \includegraphics[width=\textwidth]{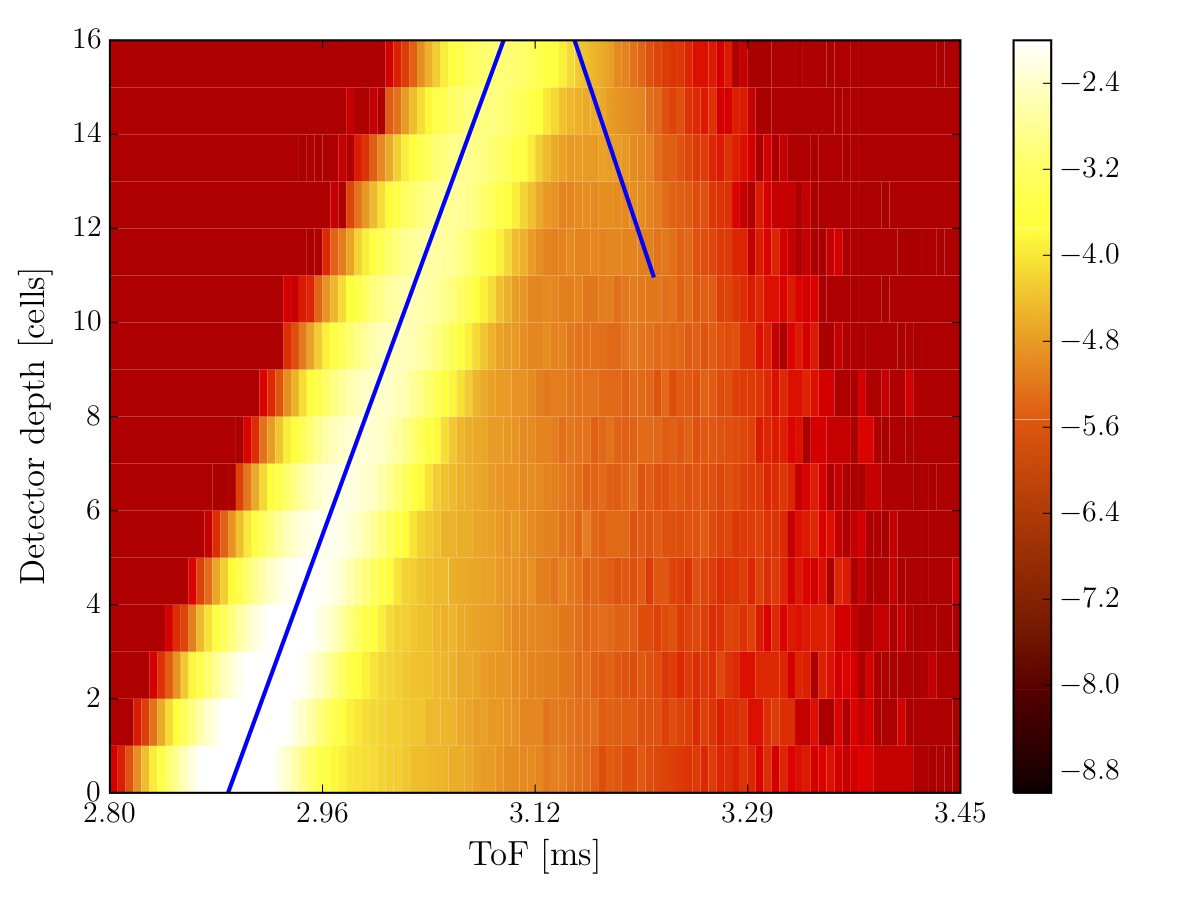}
    \caption{Simulated ToF at 4.6~\AA. \\ E$_{ini}$: Gaussian \label{2D_s_4.6_gauss}}
  \end{subfigure}
  \begin{subfigure}[b]{0.32\textwidth}
    \includegraphics[width=\textwidth]{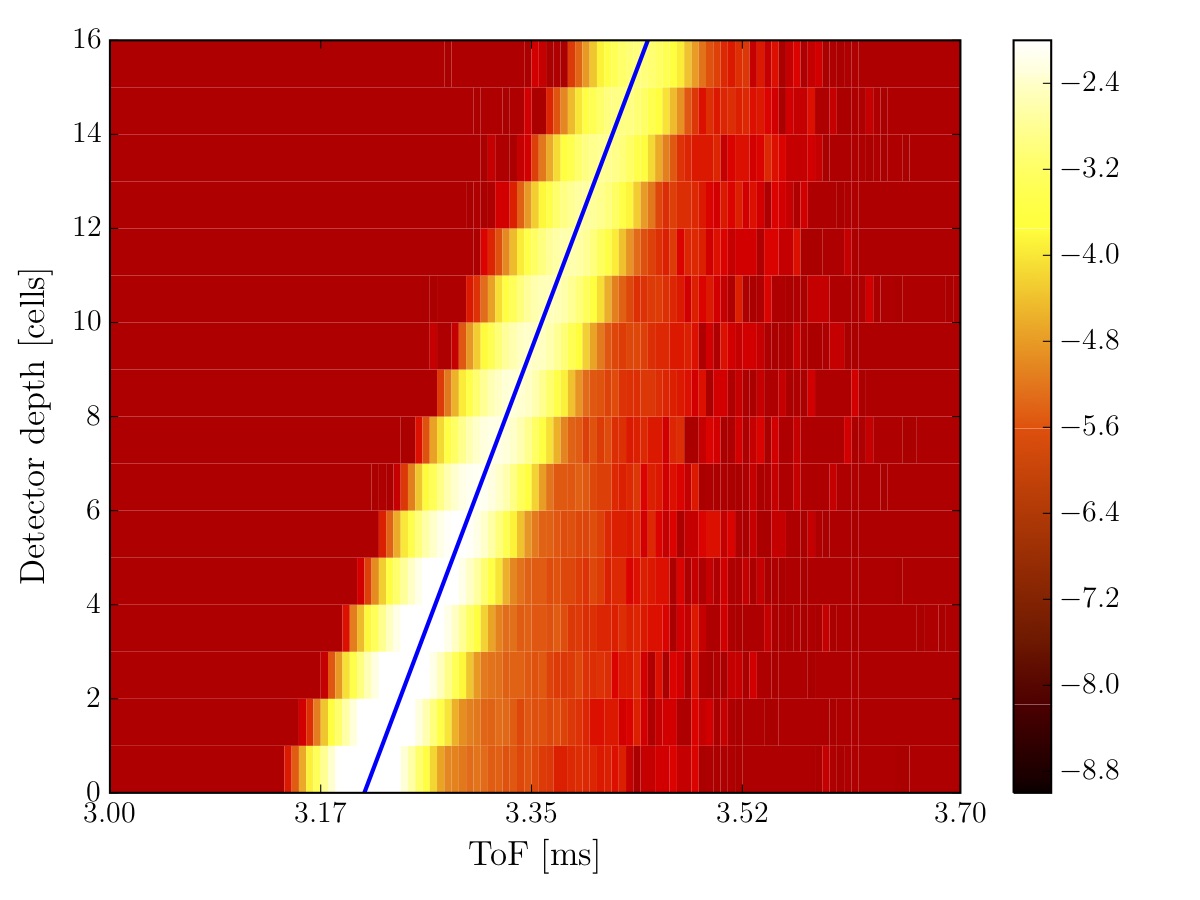}
    \caption{Simulated ToF at 5.1~\AA. \\ E$_{ini}$: Gaussian \label{2D_s_5.1_gauss}}
  \end{subfigure}

  \caption{Time-of-Flight spectra as a function of the detection depth. Results of measurement at the IN6 experiment (\Cref{2D_m_4.1,2D_m_4.6,2D_m_5.1}, measured data taken from~\cite{khaplanov2014}) and Geant4 simulation with mono-energetic~(\Cref{2D_s_4.1_mono,2D_s_4.6_mono,2D_s_5.1_mono}) and Gaussian~(\Cref{2D_s_4.1_gauss,2D_s_4.6_gauss,2D_s_5.1_gauss}) initial energy distributions. Time-of-Flight measured from sample position. (The 3 black lines in Figure~\ref{2D_m_4.1} are given by pixels with 0 counts due to low statistics.  \label{in6_2d}} %%dead channels.) \label{in6_2d}}
\end{sidewaysfigure}
%%TMP

%Also the ToF-spectra was performed and compared to the measured ones as it is shown for 4.1~\AA \ in Figure~\ref{in6_tof}. The simulation was run with Gaussian initial energy distribution, where the sigma of the distribution was estimated to fit typical instrument energy resolution and as well the measured data. The results are presented in a relative time scale. Also, the IN6 Multi-Grid demonstrator has an $\alpha$-background, sourcing from the uranium and thorium content of the non-purified aluminium of the grids. This was estimated as a continuous, flat background, and added to to the simulated ToF-spectra, in order to obtain a better comparison with the measured results.

The simulated ToF spectra are quantitatively compared with
the measured ones for all three wavelengths. The simulations are
produced with the same Gaussian initial energy distributions that were
previously applied for the 2D ToF-depth studies. The standard deviations %%sigmas
of the
distributions are estimated to fit both the typical instrument energy
resolution and the measured ToF data. In Figure~\ref{in6_tof} the
measured and simulated ToF spectra are presented in a
relative time scale. The IN6 Multi-Grid demonstrator has a
considerable $\alpha$-background~\cite{birch2015a}, coming from the
uranium and thorium content of the non-purified aluminium of the
grids. This background is random and evenly distributed in
time. Therefore, updated simulated spectra are reproduced for all
wavelengths, where a %%posterior
subsequent background correction is applied. This
is performed with a continuous, flat time-constant background added to
the simulated ToF spectra, in order to obtain a better
comparison with the measured results. The background is estimated to
fit the average measured background. In the case of 5.1~\AA, the
background is not entirely flat, which is presumably caused by
additional effects of the measurement setup and the instrument.
%DE For example according to the resolution of the Fermi-chopper would give a better description of the tails of the Gaussian (KK: where is the subject of the sentence?).
%%As an example with considering the Fermi-chopper resolution~\cite{peters2017} in the model, it would give a better description of the tails of the Gaussian ToF peaks.
As an example, adding the resolution~\cite{peters2017} of the Fermi-chopper in the model would give a better description of the tails of the Gaussian ToF peaks.
%DE (\textcolor{red}{askref from Phil})
Due to lack of additional information %%these effects are not possible to estimate.
it is impossible to estimate these effects.
%DE Due to lack of additional information on detector installatinon and measurement setup, these effects are not possible to estimate.
%% 
%% \begin{figure}[ht!]
%%   \centering

%%   \begin{subfigure}[b]{\textwidth}
%%     \includegraphics[width=\textwidth]{mg_plots/sim_meas_nm_4p6.eps}
%%     \caption{\label{in6_tof_4p6}}    
%%   \end{subfigure}

%%   \begin{subfigure}[b]{0.49\textwidth}
%%     \includegraphics[width=\textwidth]{mg_plots/sim_meas_nm_4p1.eps}
%%     \caption{\label{in6_tof_4p1}}
%%   \end{subfigure}
%%   %%
%%   \begin{subfigure}[b]{0.49\textwidth}
%%     \includegraphics[width=\textwidth]{mg_plots/sim_meas_nm_5p1.eps}
%%     \caption{\label{in6_tof_5p1}}
%%   \end{subfigure}
%
%%\FloatBarrier  %%new 
\begin{sidewaysfigure}[ht!]
  \centering

  \begin{subfigure}[b]{0.49\textwidth}
    \includegraphics[width=90mm]{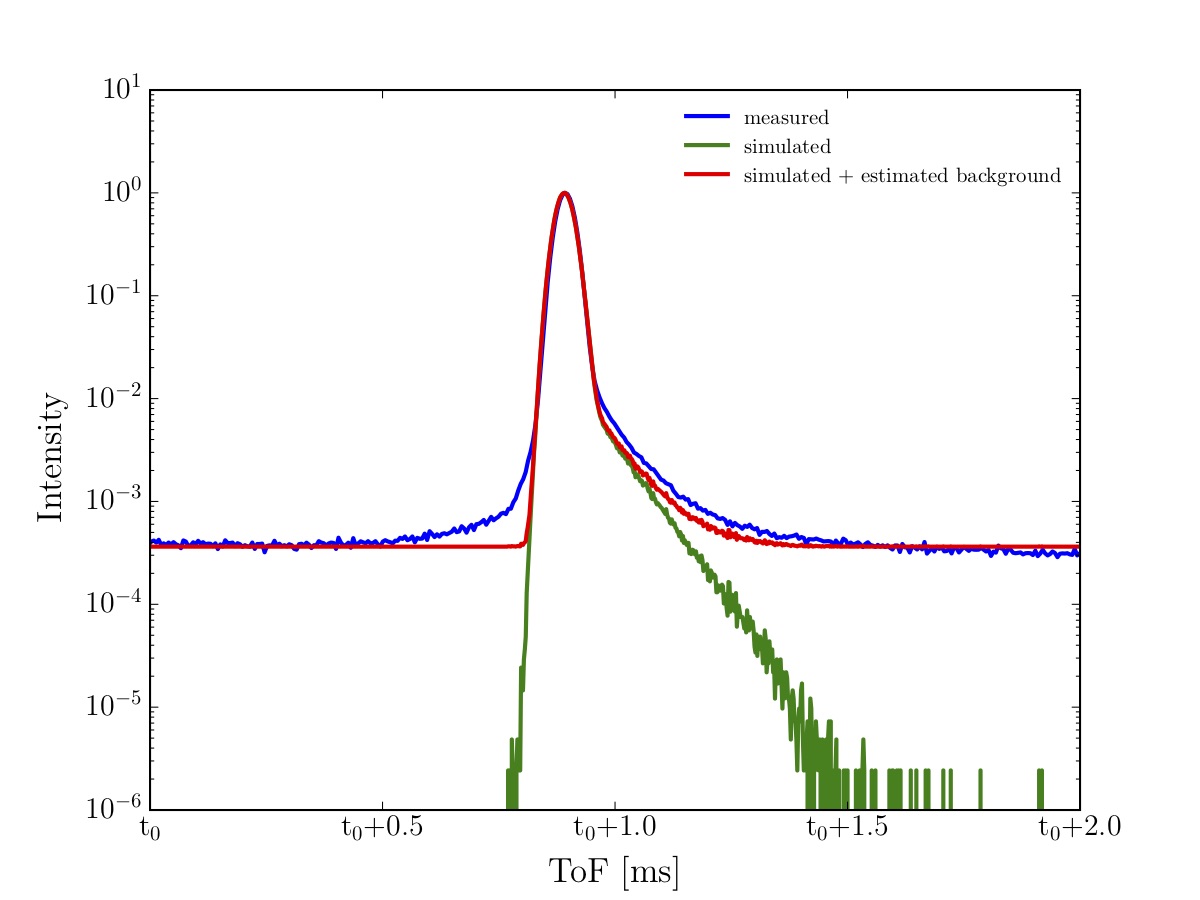}
    \caption{Simulated ToF at 4.6~\AA. \label{in6_tof_4p6}}    
  \end{subfigure}
  ~
  \begin{subfigure}[b]{0.49\textwidth}
    \includegraphics[width=90mm]{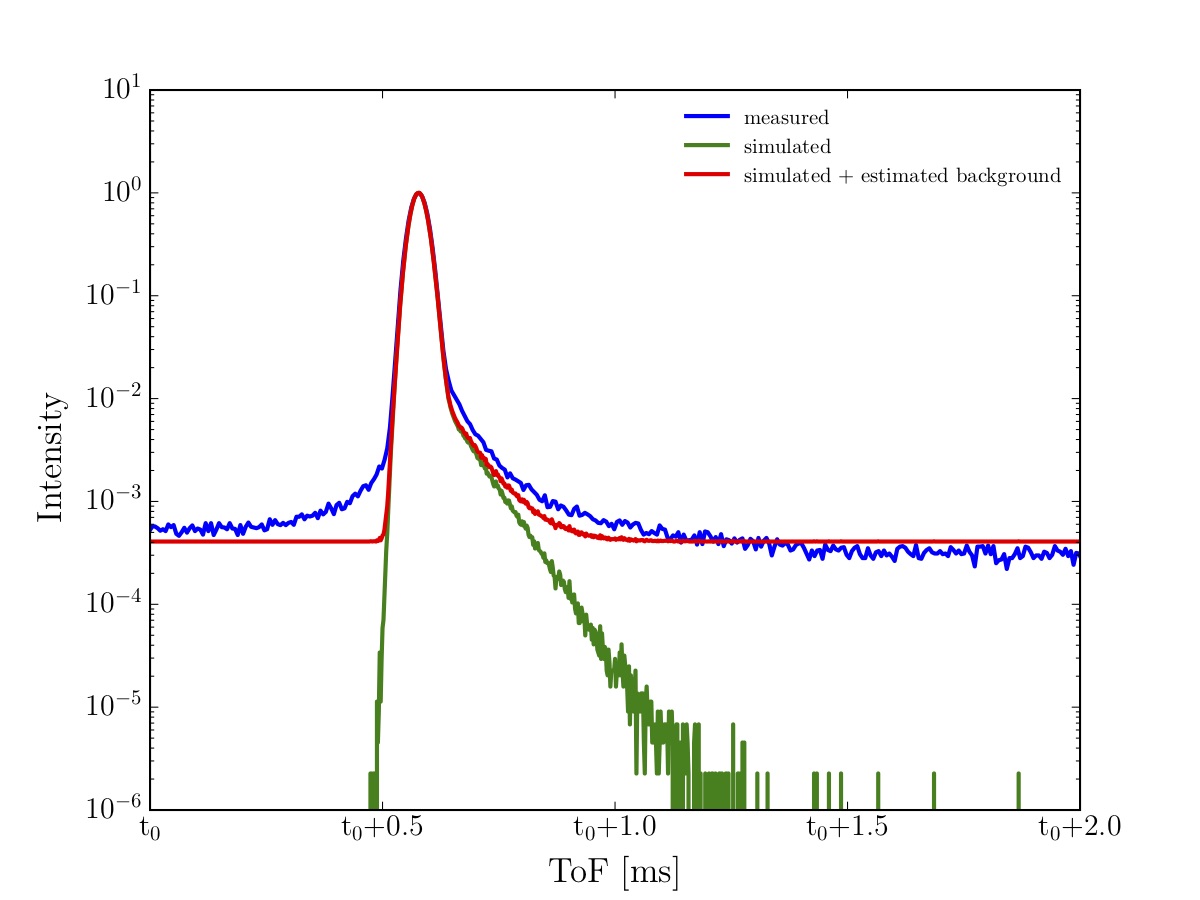}
    \caption{Simulated ToF at 4.1~\AA. \label{in6_tof_4p1}}
  \end{subfigure}
  
  \begin{subfigure}[b]{0.49\textwidth}
    \includegraphics[width=90mm]{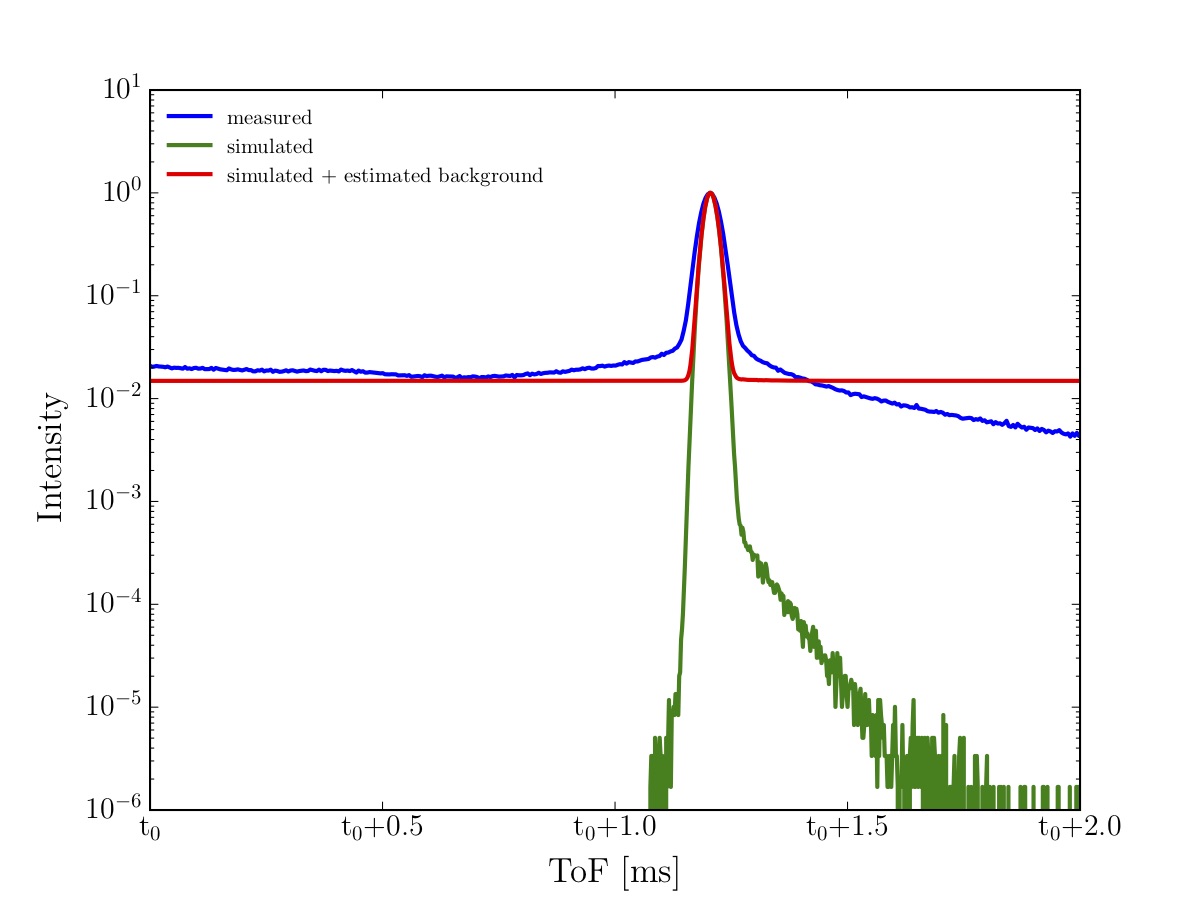}
    \caption{Simulated ToF at 5.1~\AA. \label{in6_tof_5p1}}
  \end{subfigure}
  
  \caption{Comparison of measured and simulated ToF spectra with and without $\alpha$-background correction at 4.1~\AA~(\ref{in6_tof_4p1}), 4.6~\AA~(\ref{in6_tof_4p6}) and 5.1~\AA~(\ref{in6_tof_5p1}) in relative time-scale. Intensity of Time-of-Flight spectra is given as number of counts normalised to maximum. Measured data is taken from~\cite{khaplanov2014}. $\rm t_0$ relates to the incidence of the neutron pulse on the sample position. \label{in6_tof} }
    %%release from sample position.\label{in6_tof} }
\end{sidewaysfigure}
%%TMP

%In Figure~\ref{in6_tof} it is shown that the measured and simulated ToF-peaks fit, and also that with the correction for the continuous background, the right-hand-side decay of the ToF-spectrum was also reproduced in characteristics, with only a small discrepancy in the values.
Figure~\ref{in6_tof} demonstrates that the measured and simulated
ToF peaks %%fit
agree at all the studied wavelengths. Moreover, by
applying a correction of a continuous background, the right-hand-side
decrease of the ToF spectrum is also reproduced 
%DE characteristics (KK: quantitatively??), with only a small discrepancy in the values at 4.1 and 4.6~\AA. 
 quantitatively, with only a small discrepancy in the values at 4.1 and 4.6~\AA. 

These analysis results of the IN6 model and data serve as quantitative
validation of the Multi-Grid simulation. 

The now validated model is applicable to general Multi-Grid irradiation setups, %%with more complex irradiation setups
like the CNCS demonstrator test, for identifying detector and instrument background effects.

%DE
%% (KK: pass to the next
%% section?, now you can go on and identify more complicated detector and
%% instrument background effects?)

%% \begin {enumerate}
%% \item 
%% \item CNCS:
%%   \begin {enumerate}
%%   \item Measured and simulated energy spectra
%%   \item Energy trasfer
%%   \item (ToF?)
%%   \item Relevant detector and instrument effects? 
%%   \end {enumerate}
%% \item IN6
%%    \begin {enumerate}
%%   \item Energy trasfer  (?) - let's see if we have measured  data...
%%   \item(ToF?)
%%   \item ToF spectra as a function of depth in detector [Ref Anton's paper on IN6] - Still need for goof FoM 4 qualitative comparison
%%   \end {enumerate}
%% \end {enumerate}

%DE
%% \subsection{Model of demonstrator test on CNCS instrument at SNS KK: a
%% more concise title, think of turning it into a section as with IN6} \label{cncsmod}%The model of the Multi-Grid prototype used at the IN6 experiment is a detector of 6 columns, with 16 grids in each. For this detector

\section{Model of demonstrator test on CNCS instrument at SNS} \label{cncsmod}
%DE
%% A two-column Multi-Grid prototype is tested~\cite{khaplanov2017} at
%% the CNCS instrument. The detector columns consist of 2~x~48 grids, 4~x~17 cells in each grid, with 1~mm Gd$_{2}$O$_{3}$ shielding on the rear end of the grids, as shown in Figure~\ref{ccncs}~and~Figure~\ref{tcncs}. A 2~mm thick %%boron-carbide-based shielding layer
%% MirroBor~\cite{mirrobor} rubber layer with 80 mass~\% B$_{4}$C
%% content withnatural boron %DE 80~\% enriched in $^{11}$B (KK: do you mean natural?)   
%% is also inserted between the columns to reduce cross-scattering. The detector is filled with 80/20 volume~\% Ar/CO$_{2}$ at %%$\rm p = 1~bar$ pressure
%% nominal room temperature and pressure, and placed in an aluminium
%% vessel with the same gas-filling. The distance between the sample position and the front face of the
%% detector vessel was 333.5~cm. The Geant4 model of the detector was
%% built with the same parameters, although the bottom and rear end
%% electronics were neglected as well as the external
%% boron-carbide-shielding of the vessel.(KK: again boring :-) Pick only what
%% is necessary or different to the previous module. careful with the gas
%% composition, was it the same inside and outside the detector? Or
%% 98/2??, fix tense)

A two-column Multi-Grid prototype is tested~\cite{khaplanov2017} at
the CNCS instrument. The detector columns consist of 2~x~48 grids, with 1~mm Gd$_{2}$O$_{3}$ shielding on the rear end of the grids, and a 2~mm thick %%boron-carbide-based shielding layer
MirroBor~\cite{mirrobor} rubber layer with 80 mass~\% natural B$_{4}$C
content %DE 80~\% enriched in $^{11}$B (KK: do you mean natural?)   
is also inserted between the columns to reduce cross-scattering.

The columns are placed in an aluminium vessel, as shown in Figure~\ref{ccncs}~and~Figure~\ref{tcncs}, and the whole  detector volume is filled with \textit{counting gas}: Ar/CO$_{2}$ (80/20 by volume) at 
nominal room temperature and pressure.
The Geant4 model of the detector was built with the same parameters.

In this model some of the instrument components are also present.
The measurement chamber is filled with \textit{tank gas}: Ar/CO$_{2}$ (98/2 by volume) at 
nominal room temperature and pressure. Tank gas is the gas in the cylindrical chamber on the flight path between the sample and the detector.

A simplified model of the sample environment
is also implemented. It consists of a double-wall aluminium cylinder with
radii of 10 and 12~cm and a 2~mm wall-thickness, representing the
cryostat, and a 0.5~mm thick aluminium window with 74~cm radius (see Figure~\ref{beam_sampenv_cncs}), representing the barrier between air and tank gas.
%%REVISED
In addition a 2$^\circ$ collimator is involved, placed between the cryostat and the aluminium window. The collimator is built of 136 pieces of 1~m high and 10~cm long stainless steel blades with $\rm 2 \times 10\ \mu$m Gd$_{2}$O$_{3}$ painting.

%%REVISED
A significant effort has been made to understand and reduce the background in the Multi-Grid and other solid boron converter based detectors. As a part of this, the $\alpha$-, $\gamma$- and fast neutron background components have been studied and reduced, as described in \cite{birch2015a}, \cite{khaplanov2013a} and \cite{mauri2018}, respectively. These background components are also omitted from the simulation, as the remnant background is negligible in comparison with the implemented instrument-related background sources \cite{khaplanov2017}.

A series of tests are performed and published with this
measurement setup, %DE (KK: I know only 1),
and the high statistics results
%%REVISED
%%with a vanadium sample~\cite{khaplanov2017} are selected for simulating.
with a vanadium sample~\cite{khaplanov2017} at 1.0, 3.678 and 3.807~meV (i.e. 9.04, 4.72 and 4.64~\AA, respectively) are selected for simulating. 
%DE for the validation of the model (KK: I thought you validated it already. Maybe stop repeating it?).
In order to identify %DE eliminate (KK: or identify?, distinguish)
the scattered background components, the simulations are
repeated with multiple geometry configurations, e.g.\,with and without
sample environment or detector vessel, as well as with multiple
neutron generators, e.g.\, a targeted beam irradiating the entire
detector surface or
%DE an isotropic source (KK: careful with this word, you keep using it instead of 4pi or large solid angle, your beams are always isotropic)
a 4$\pi$-source, all with mono-energetic and Gaussian initial neutron energy distributions. 
The $\sigma$ of the Gaussian distribution is chosen %%arbitrarily
as 0.006~meV for the 1.0 and 0.030~meV for the 3.678 and 3.807~meV incident neutron energies, respectively, to fit the measured data, considering the known 1~\% resolution of the CNCS instrument.

The complete model of detector \&  sample environment was checked with the simulation of the directly measured ToF and flight distance data.
%
%DE
%% During the measurement the Time-of-Flight, flight-distance and energy
%% transfer were determined, so the same quantities were used for the
%% validation. (KK; rephrase, no more validation pleaaaaseeee :-), write
%% a pass to the next section or nothing. Let the first sentence of the
%% next subsection guide the reader.)
%% %%PHS, dE/E??
%
\FloatBarrier
\begin{figure}[ht!]
  \centering
  
  \begin{subfigure}[b]{0.18\textwidth}
    \includegraphics[width=\textwidth]{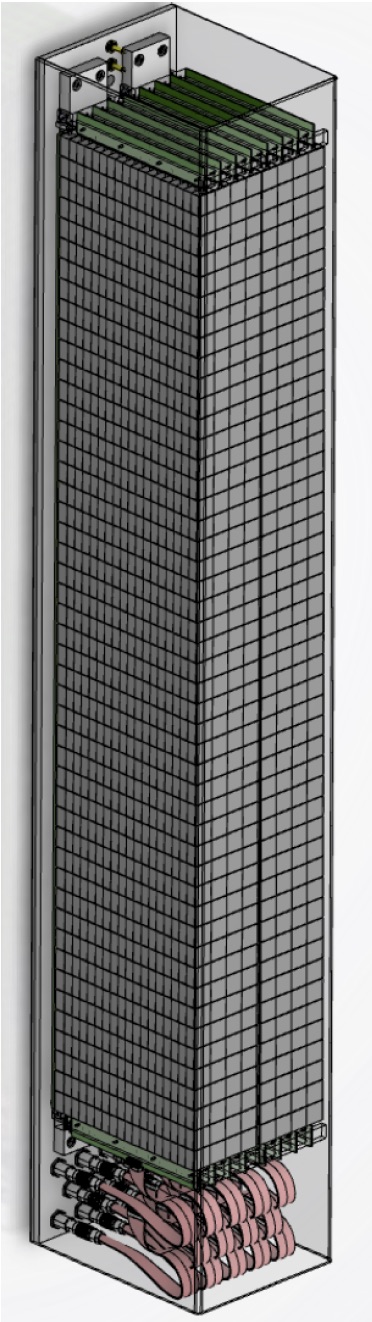}
    \caption{\label{ccncs_draw}}
  \end{subfigure}
  \begin{subfigure}[b]{0.18\textwidth}
    \includegraphics[width=\textwidth]{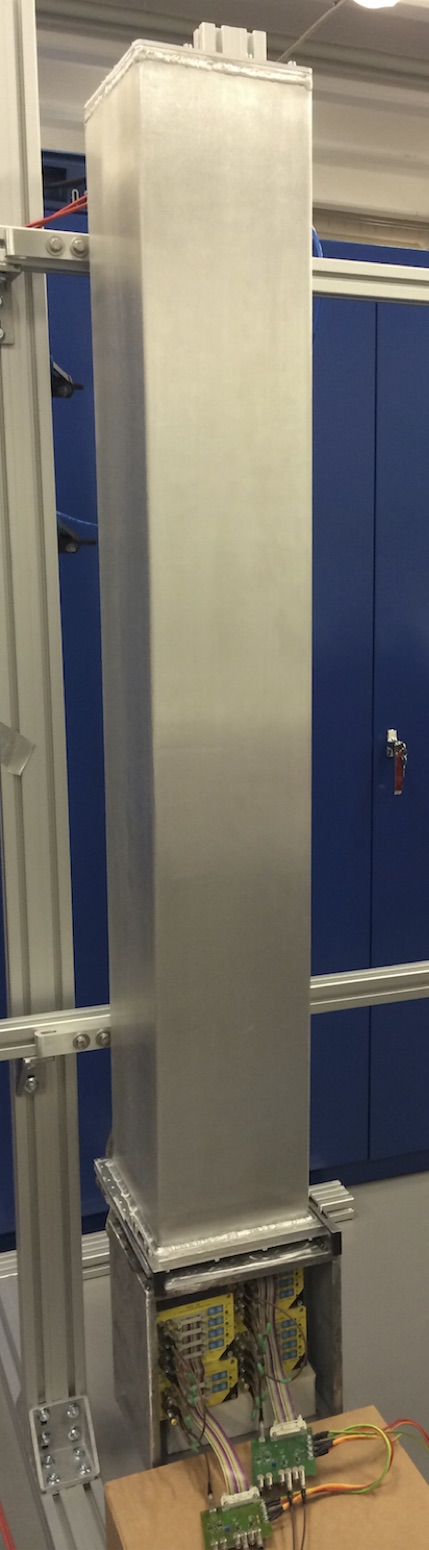}
    \caption{\label{ccncs_real}}
  \end{subfigure}
  \begin{subfigure}[b]{0.2\textwidth}
    \includegraphics[width=\textwidth]{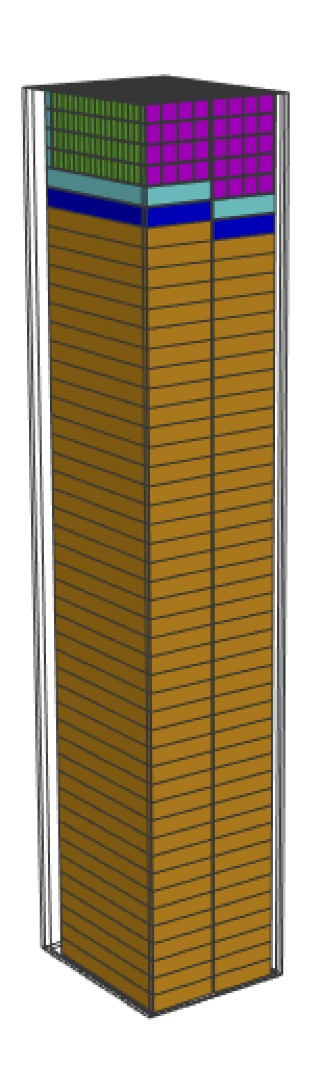}
    \caption{\label{ccncs_g4}}
  \end{subfigure}
  
  \caption{The CNCS demonstrator: technical drawing in CATIA~V6~\cite{catia6}~(\ref{ccncs_draw}, source of plot:~\cite{khaplanov2017}), built prototype (\ref{ccncs_real}, source of plot:~\cite{khaplanov2017}) and Geant4 model (\ref{ccncs_g4}).  \label{ccncs}}
\end{figure}
%new
\begin{figure}[ht!]
  \centering

  %% \begin{subfigure}[b]{0.35\textwidth}
  %%   \includegraphics[width=\textwidth]{mg_plots/CNCStop_draw.jpg}
  %%   \caption{\label{tcncs_draw}}
  %% \end{subfigure}
  %% \begin{subfigure}[b]{0.37\textwidth}
  %%   \includegraphics[width=\textwidth]{mg_plots/CNCStop_G4.jpg}
  %%   \caption{\label{tcncs_g4}}
  %% \end{subfigure}
  \begin{subfigure}[b]{0.35\textwidth}
    \includegraphics[width=\textwidth]{Grid_photo.jpg}
    \caption{\label{tcncs_draw}}
  \end{subfigure}
  \begin{subfigure}[b]{0.47\textwidth}
    \includegraphics[width=\textwidth]{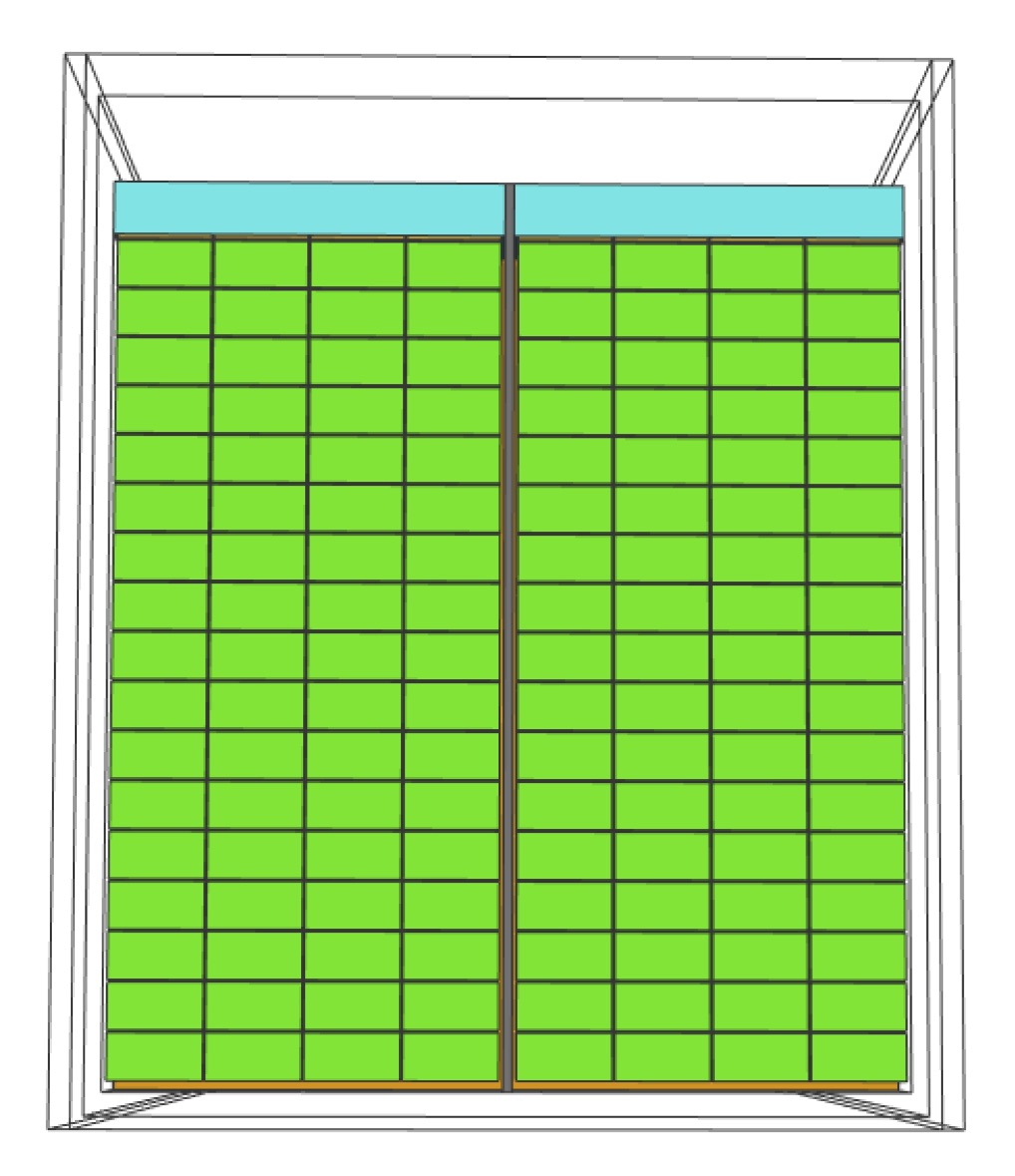}
    \caption{\label{tcncs_g4}}
  \end{subfigure}

  %% \caption{Inner layout of the CNCS demonstrator: technical drawing (\ref{tcncs_draw}, source of plot:~\cite{khaplanov2017}) and Geant4 model (\ref{tcncs_g4}).  \label{tcncs}}
  \caption{Layout of the CNCS demonstrator: single grid~(\ref{tcncs_draw}) and Geant4 model~(\ref{tcncs_g4}). \label{tcncs}}
    % \textcolor{blue}{KK: doyou need these figures? Perhaps a better full geometry snapshot?)}  \label{tcncs}}
\end{figure}
%new
%
\begin{figure}[ht!]
  \centering
  %% \includegraphics[width=\textwidth]{mg_plots/cncs_cylinder_source_geom.jpeg}
  %% \caption{Geometry view of CNCS Geant4 model with sample environment,
  %%   cylindrical source and detector module.\label{beam_sampenv_cncs}}
  %%\includegraphics[width=\textwidth,natwidth=1872,natheight=1146]{cncs_4pi_source_geom.jpg}
  %%\includegraphics[width=\textwidth]{cncs_4pi_source_geom.eps}
  \includegraphics[width=\textwidth]{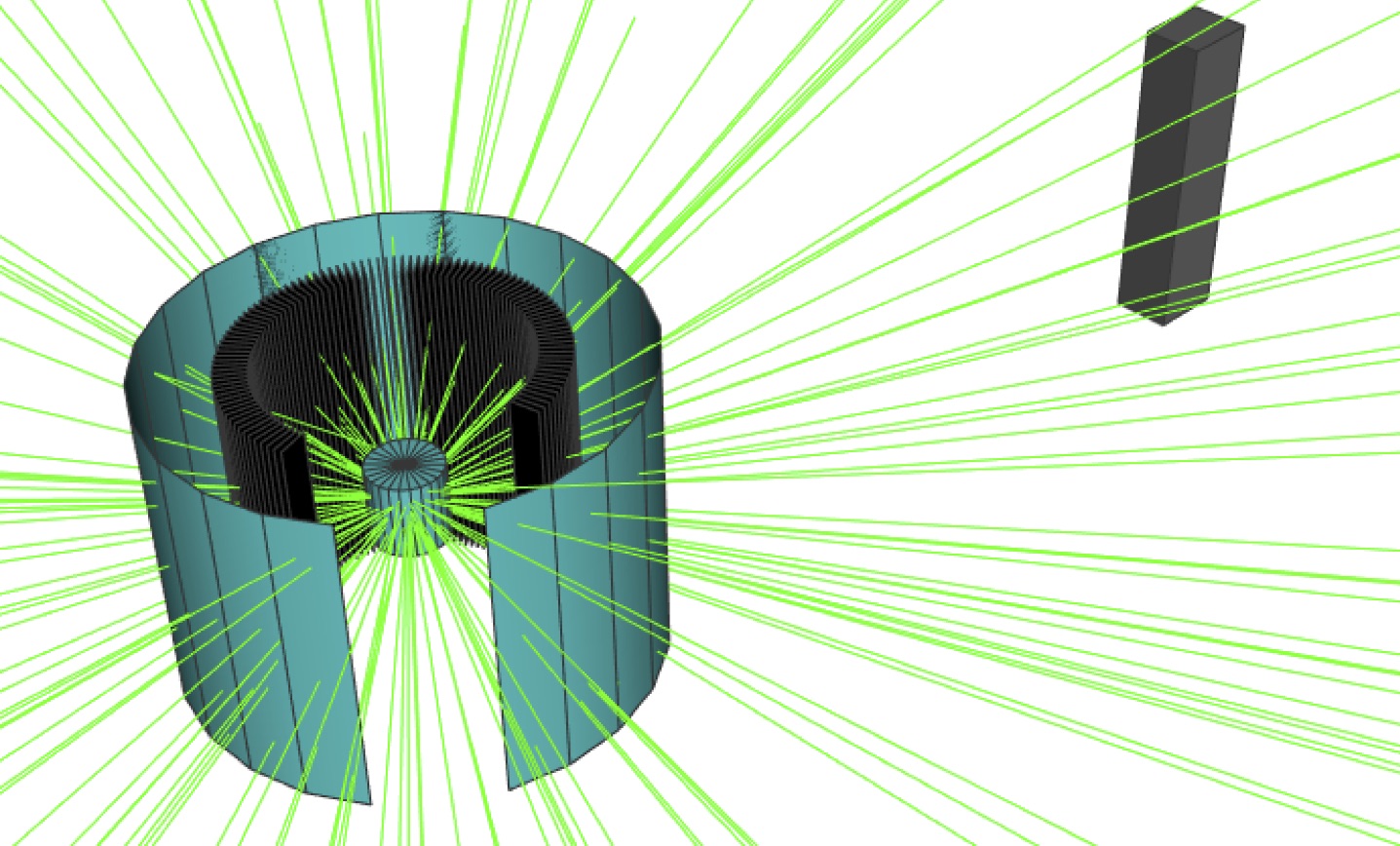} %%REVISED
  \caption{Geometry view of CNCS Geant4 model with sample environment,
    4$\pi$-source and detector module.\label{beam_sampenv_cncs}}
    %DE KK: fix caption according to a previous example.
    % Geant4 doesn't offer visualisation \label{beam_sampenv_cncs} }
\end{figure}

\subsection{Simulation results of primary measured data for CNCS demonstrator detector}\label{cncsRnToF}

%DE
%% In the case of chopper spectroscopy the information (KK: which
%% information?) is carried by the momentum- and energy transfer of the
%% scattered neutrons. These are derived from the primary measured
%% quantities: the detection coordinates (=flight distance) and the
%% Time-of-Flight, therefore these raw measured quantities were simulated
%% for validation (KK: shouldn't you have said this much earlier??). The simulations were performed %at 3.678 and 3.807~meV incident neutron energies
%% with the most realistic case of the simulation:
%% \begin{itemize}
%% \item realistic detector model in aluminium vessel
%% \item Gaussian initial neutron energy distribution
%% \item presence of simplified sample environment, including cryostat and aluminium window
%% \item presence of room temperature and pressure Ar/CO$_{2}$~ 98/2
%%   volume \% gas in tank
%%   \item KK: you have said almost all of it in the previous subsection
%%     and this time with correct ambient gas composition!
%% \end{itemize}

The directly measured quantities, the ToF and the flight distance are simulated for checking the implemented detector and instrument setup of all the afore mentioned components.

The measured and simulated ToF and flight distance spectra at 3.678 and 3.807~meV incident neutron energies, below and above the aluminium Bragg-edge, are compared in Figures~\ref{R} and \ref{tof}. 

%DE
%% The flight distance is defined as the distance from the sample
%% position to the detection coordinates (KK: has to come earlier). The
%% simulated detection coordinates are similarly reduced to the center of
%% the cell in which the neutron is detected, despite the higher
%% resolution of the simulation. (KK: this has to be written when you
%% introduce simulations, also where is the 100 keV threshold? say that
%% simulation is trying to reprodue the experimental conditions closely)

%% (KK: that's not what binning does, remove sentence, nobody cares about
%% your binning) A realistic binning is chosen to have
%% high enough statistics for the comparison.

As shown in Figures~\ref{R_3p678} and \ref{R_3p807}, a series of peaks
appear in both measured and simulated flight distance spectra,
relating to the geometrical cell structure of the grids.
%%REVISED
The resolution of the detector is affected by this cell structure, therefore these peaks are related to the rows of cells in the detector. %%
The peaks are visible in the first 10--15~cm of the detector, where the majority of the
neutrons are detected, therefore the statistics are the highest. The
falling tail of the spectra is determined by the neutrons detected in
the rear cells of the detector, and by the scattered neutrons, having a
longer flight distance. There is a difference in the cutoff of the two
spectra, since the last row is not read out in the measurement,
contrary to the simulation.
%new
\FloatBarrier
\begin{figure}[ht!]
  \centering

  \begin{subfigure}[b]{\textwidth}
    \includegraphics[width=90mm]{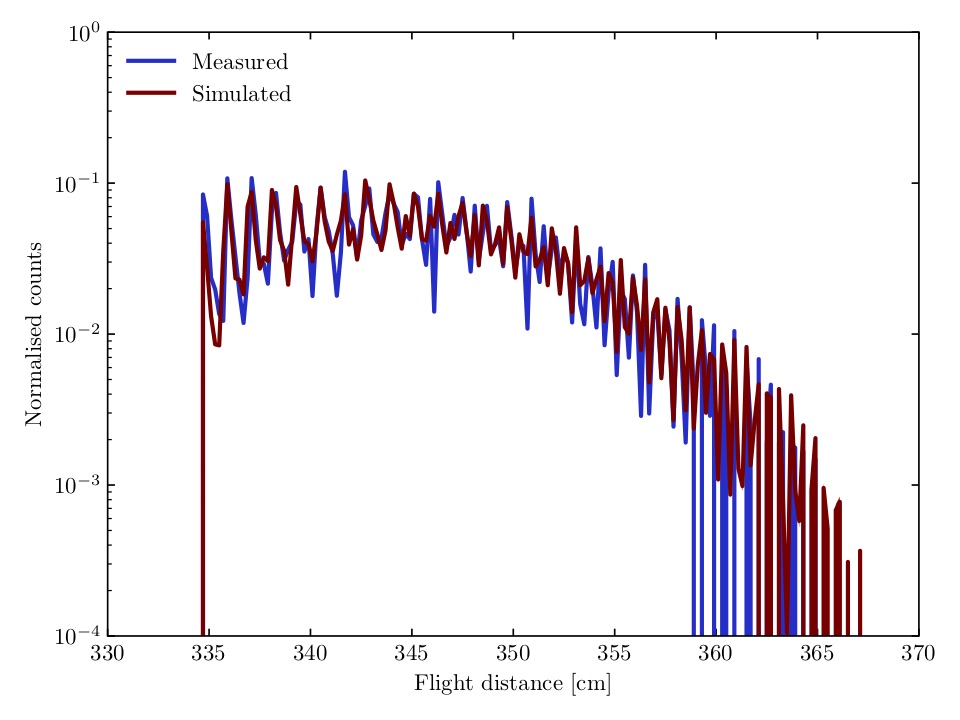} %R_c_200b
    \caption{Flight distance at 3.678~meV. \label{R_3p678}}
  \end{subfigure}
  
  \begin{subfigure}[b]{\textwidth}
    \includegraphics[width=90mm]{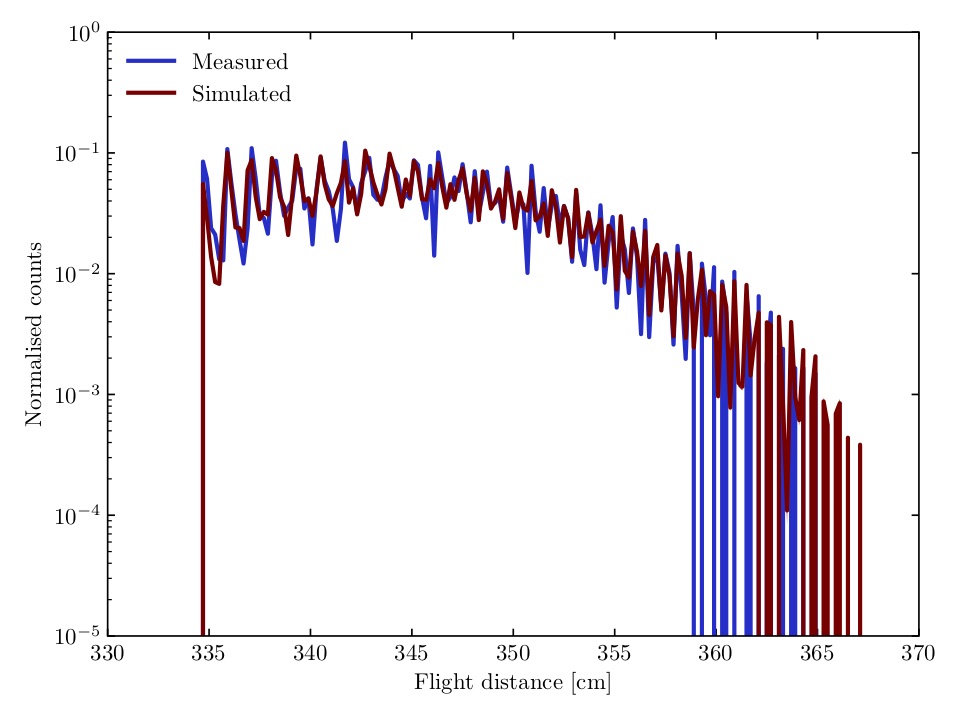}
    \caption{Flight distance at 3.807~meV. \label{R_3p807}}
  \end{subfigure}
  
  \caption{Measured and simulated flight distance spectra at 3.678~(\ref{R_3p678}) and 3.807~meV~(\ref{R_3p807}) incident neutron energies, normalised to area. \label{R}} %% Data for comparison was measured on CNCS at SNS under ID IPTS-17219. \label{R}}
\end{figure}

Both the overlaying peaks  and the characteristics of the falling
tails of the measured and simulated spectra are in good agreement at
both energies below and above the aluminium Bragg-edge.
%DE , serving as qualitative validation of the built model. (KK: again validation?)

The measured and simulated ToF spectra are compared in
Figure~\ref{tof}. %% The simulated Time-of-Flight is measured from the
%% sample position, while in the experiment it is measured from the
%% chopper \textcolor{blue}{(KK: which chopper? be specific)}.
The simulated ToF is measured from the sample position, while the experimental data are given relatively to the 16667~$\rm \mu s$ period of the SNS pulse.
%% An arbitrary shift is applied on the measured spectra to overlay them with the simulated ones. The measured and simulated Time-of-Flight peaks are at perfect agreement %within the margin of error. Add errorbars!!! TODO
%% at both energies.
%% However, there are significant discrepancies between
%% the measured and simulated backgrounds.
%% %% The source of this discrepancy
%% %% is that not all instrument related effects, e.g.\,instrument
%% %% background radiation, initial Time-of-Flight distribution of neutrons,
%% %% and some of the sample environment components, like radial collimator
%% %% are included in the simulation, since the aim of the current study
%% focusses on understanding detector effects. %the detector prototype, not the whole experimental setup.

An arbitrary shift is applied on the measured spectra to overlay them with the simulated ones. This way the measured and simulated ToF peaks are fit at both energies; the shape and the width of the peaks give good agreement.

In both spectra the measured and simulated backgrounds also reasonably agree with the presence of some discrepancies between them.
The source of these discrepancies is that not all instrument related effects are included in the simulation.
For example instrument background radiation, initial ToF distribution of neutrons,
and some of the sample environment components %%REVISED, like the radial collimator
are omitted, since the aim of the current study
focuses on understanding detector effects. However, the level of agreement of the measured and simulated backgrounds are acceptable, considering the diversity of backgrounds of the existing chopper spectrometers.

%DE
%% In essence, the agreement of the Time-of-Flight peaks both below and
%% above the aluminium Bragg-edge validates the built model both
%% qualitatively and quantitatively. (KK: you know what I want, these
%% sentences are repeated way too often.)
%
In essence, the measured and simulated ToF of elastic peaks agree well.%%DE also give good agreement.

Therefore, the now tested CNCS irradiation setup can be used for performing detailed scattered neutron background study.
%new
\FloatBarrier
\begin{figure}[ht!]
  \centering
  
  \begin{subfigure}[b]{\textwidth}
    \includegraphics[width=90mm]{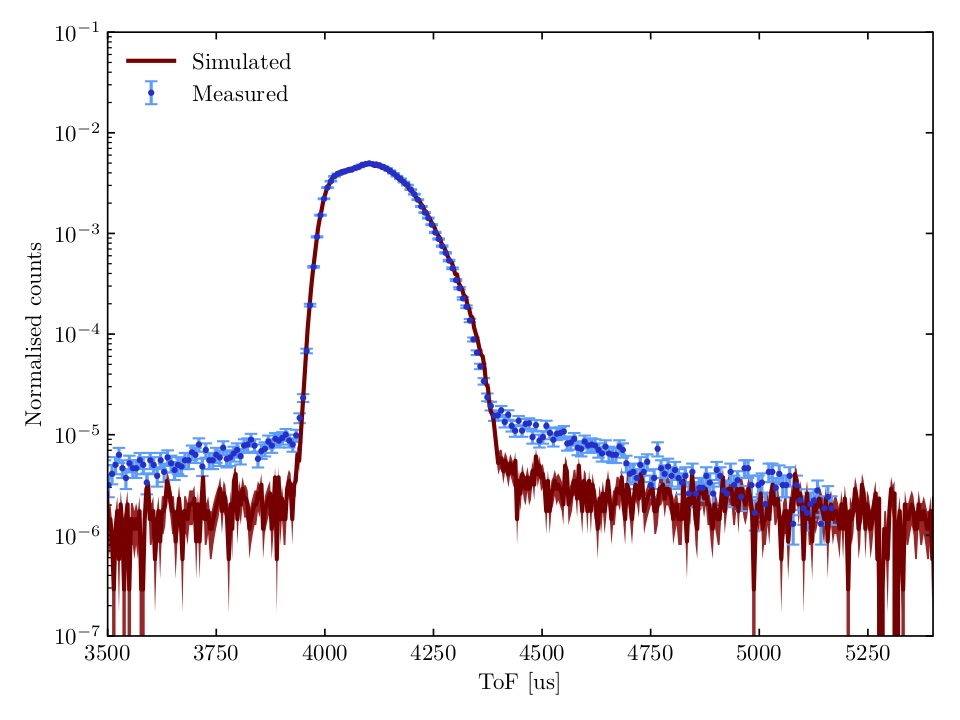}
    \caption{ToF at 3.678~meV. \label{tof_3p678}}
  \end{subfigure}
  
  \begin{subfigure}[b]{\textwidth}
    \includegraphics[width=90mm]{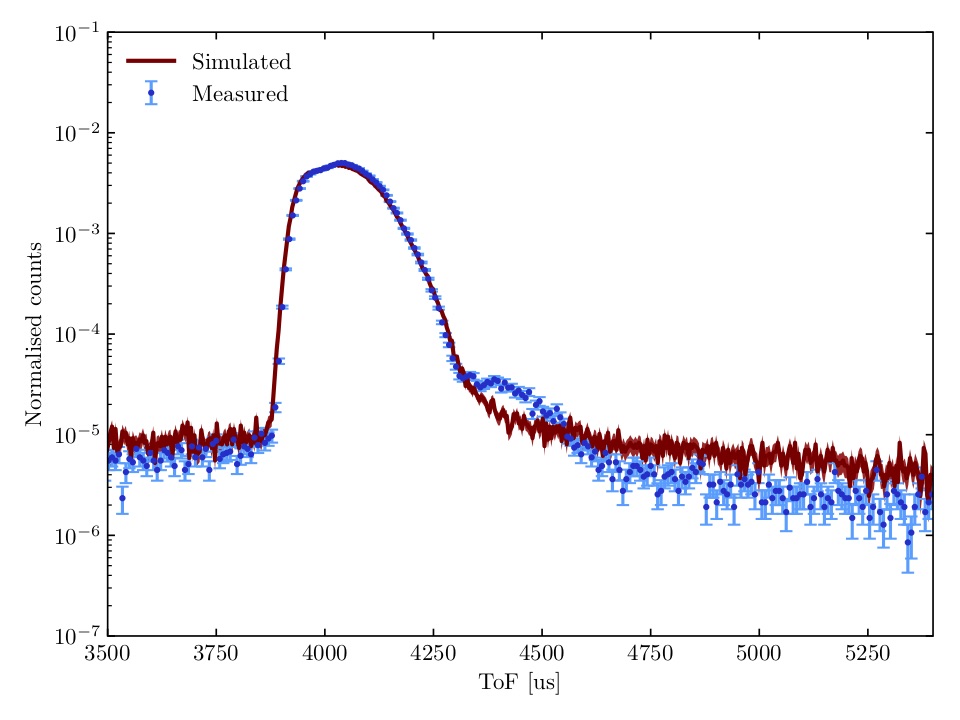}
    \caption{ToF at 3.807~meV. \label{tof_3p807}}
  \end{subfigure}
  
  \caption{Measured and simulated Time-of-Flight spectra at 3.678~(\ref{tof_3p678}) and 3.807~meV~(\ref{tof_3p807}) incident neutron energies, normalised to area. \label{tof}} %%  Data for comparison was measured on CNCS at SNS under ID IPTS-17219. \label{tof}}
\end{figure}

%% \subsection{Simulation results of measured (KK: it is derived, not measured) energy transfer and background differentiating)}%DE elimination) 
\subsection{Energy transfer and background differentiation from measurement and simulation}\label{cncsEtrf} %DE elimination)
  %DE(KK: is that the right word?) for CNCS demonstrator detector}\label{cncsEtrf}

%The final goal of detector modelling is to develope and validate a Multi-Grid detector model that can be used as part of the optimisation process of the detector. The main aim is to increase the SBR of the detector by reducing the scattered neutron background with the application of optimised detector shielding. For this the detailed understanding of the neutron scattering sources is essential.

Measured and simulated energy transfer spectra are compared as part of
the validation process of the implemented Multi-Grid detector
model. Simulations are also performed with different geometries, with
the additional aim of identifying and distinguishing the sources of
neutron scattering. The energy transfer is defined as $E_{trf} =
E_{initial} - E_{final}$, therefore the elastic peak appears centred around
0~meV, while the negative side represents the neutrons detected with
energy gain and the positive side represents the neutrons with energy
loss in comparison with the initial energy. For this study the
detector background is defined as all neutron events in the energy
transfer spectrum outside of the elastic peak. Since the peaks are sharp and well-identifiable, the peak borders are estimated %%manually, by the eye.
visually. The background is always given normalised to the peak: % $background = (Total \ counts - Counts \ in \ peak) / Counts \ in \ peak$.
%DE \textcolor{blue}{(KK personal note: need to think how to better express this, perhaps the
%DE normalization has to be mentioned later)}

\begin{equation}\label{eq:bg} 
  background \ fraction = \frac{Total \ counts - Counts \ in \ peak}{Counts \ in \ peak}. \\
\end{equation}

The simulations are performed in the 1.0--8.0~meV incident neutron energy range. %%The measured data was compared with the most realistic case of the simulation, involving the same geometrical and instrumental component as in the case of the Time-of-Flight and flight distance comparisons (see Section~\ref{cncsRnToF}).
The measured and simulated energy transfer spectra at 1.0, 3.678 and
3.897~meV incident neutron energy, below and above the aluminium Bragg
edge are presented in \Cref{E_1p000_v,E_3p678_v,E_3p807_v} respectively.
Simulations are repeated adding one-by-one the geometrical and instrumental components (see Section~\ref{cncsmod}) to the simulation. The spectra are compared in \Cref{E_1p000_bg,E_3p678_bg,E_3p807_bg}, while the obtained scattered neutron background data are given in Table~\ref{tab:bg_E_trf}. ``Bare detector grids'' means two columns of grids, without the aluminium vessel.

In the energy transfer spectrum of the bare grids the elastic peak is
mono-energetic at 0~meV and an asymmetric scattered neutron background
also appears. The source of the background on the negative side is %%are
the neutrons that gained energy via inelastic scattering.
%DE (KK: you mean inelastic?? are you referring to the derived energy?). %%The longer background on the positive side appear as neutrons with energy loss. However,
%One part in the higher and wider background on the positive side is given by the neutrons that lost energy in inelastic scattering. This is a less probable than the enrgy gain, since the cold neutrons were thermalising in the room temperature environment. However,
The major source of higher and broader background on the positive side %%are
is the contribution of the elastically scattered neutrons. Since the flight distance is
calculated from the detection coordinates, assuming the shortest
%%distance
path between the detection point and the sample, but the
ToF measured represents the entire neutron path, in the
case of elastic scattering a longer ToF is combined with a
shorter flight distance, resulting to the registration of an
effectively slower neutron. These neutrons have an apparent energy
loss, and cause the asymmetrical, high intensity background. In
Figure~\ref{E_3p807_bg} a fine structure of peaks also appears near
the elastic peak on the positive side: this peak relates to the grid
structure, the coherent scattering between the aluminium
blades. Therefore this effect appears only above the aluminium Bragg-edge.

A similar spectrum is obtained with the complete detector model inside
the vessel. The scattered neutron background increases on both sides
with respect to the one of the bare grids.%~(see Table~\ref{tab:bg_E_trf}).%DE (\textcolor{red}{is this numerical interpretation fine? What to conclude from numbers?})

%% As a next step, a Gaussian initial neutron energy is added to the
%% neutron generator. The $\sigma$ of the distribution is chosen
%% arbitrarily as 0.030~meV to fit the measured data, considering the
%% known 1~\% resolution of the CNCS instrument. As it appears in
%% Figure~\ref{E_3p678_bg} and \ref{E_3p807_bg}, the initial energy
%% distribution defines the shape of the elastic peak, while its impact on the background is negligible. The realistic Gaussian distribution only affects the background by the increased peak width. It is also revealed that the coherent scattering effects of the blades are hidden in the case of realistic incident neutron energy distributions.

The effect of a Gaussian initial neutron energy distribution appears in \Cref{E_1p000_bg,E_3p678_bg,E_3p807_bg}; the initial energy distribution defines the shape of the elastic peak, while its impact on the background is negligible. The realistic Gaussian distribution only affects the background by the increased peak width. It is also %%revealed
apparent that the coherent scattering effects of the blades are hidden in the case of realistic incident neutron energy distributions.

%% Including the Ar/CO$_{2}$ gas to the measurement tank (KK: pick one
%% name to call it for the entire paper)
Including the tank gas and components of the sample environment, a continuous, flat scattered neutron background appears in the spectra. In all cases, the asymmetric detector background has a comparable shape, appearing as a shoulder on the side of the elastic peak. While at 1.0~meV~(Figure~\ref{E_1p000_bg}) and 3.678~meV~(Figure~\ref{E_3p678_bg}) the background is coming from the tank gas and the sample environment are comparable, at 3.807~meV~(Figure~\ref{E_3p807_bg}) the aluminium sample environment becomes the dominant source of background, significantly increasing the background.
%%REVISED
This background is slightly reduced by collimator, eliminating the scattered fraction of the cryostat and the backwall of the aluminium window. However, the sample environment remains the main background source above the Bragg-edge even in the presence of the collimator.%%

The measured data are compared with the most realistic case of the simulation, including all the afore described geometrical and instrumental components as in the case of the ToF and flight distance comparisons (see Section~\ref{cncsRnToF}). %% As it is shown in \Cref{E_1p000_v,E_3p678_v,E_3p807_v}, the elastic peak is qualitatively and quantitatively reproduced by the simulation in all cases. The shape of background is also reproduced. The quantitative discrepancies in the background are derived from the same sources as for the primary quantities. (see Section~\ref{cncsRnToF}).

As it is shown in \Cref{E_1p000_v,E_3p678_v,E_3p807_v}, the energy transfer spectra are reproduced by the simulation in all cases. In the case of 1.0 and 3.678~meV incident neutrons, below the aluminium Bragg-edge, the simulated background underestimates the measured one on both sides of the elastic peak. The discrepancy is about 80\%. In the case of 3.807~meV incident neutrons, above the aluminium Bragg-edge, the simulated background slightly overestimates the measured one. The discrepancy is about 20\% on the negative and 5\% on the positive side of the elastic peak. The discrepancies in the background are
%%derived from the same sources
attributed to the same reasons 
as for the primary quantities (see Section~\ref{cncsRnToF}). It also has to be mentioned that the two bumps at 0.25 and 0.5~meV only appear in the measured energy transfer. This effect is related to the instrument, as it also appears in the response of local $^{3}$He-tubes. Its independence from the presence of the Multi-Grid detector is satisfactorily verified elsewhere.

In essence the measured and simulated elastic peaks agree well and the backgrounds reasonably agree at all energies.
\FloatBarrier %new
%(\textcolor{red}{mention 3.807 bumps?})
%
%\FloatBarrier
%
%% \begin{figure}[ht!]
%%   \centering
%%
%%   \begin{subfigure}[b]{0.85\textwidth}
%%     \includegraphics[width=\textwidth]{mg_plots/3p678_E_trf_nat_sim_m.eps}
%%     \caption{Energy transfer at 3.678~meV. \label{E_3p678_v}}
%%   \end{subfigure}
%%  
%%   \begin{subfigure}[b]{0.85\textwidth}
%%     \includegraphics[width=\textwidth]{mg_plots/3p678_E_trf_nat_sim.eps}
%%     \caption{Energy transfer at 3.678~meV. \label{E_3p678_bg}}
%%   \end{subfigure}
%%
\begin{figure}[ht!]
  \centering
  
  \begin{subfigure}[b]{\textwidth}
    \includegraphics[width=90mm]{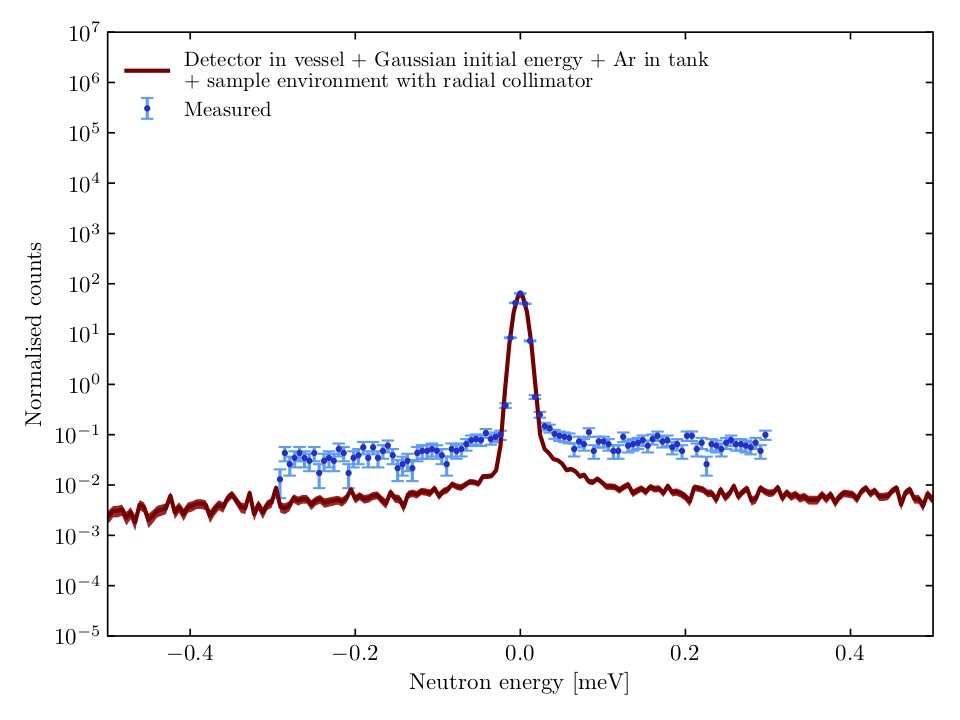}
    \caption{Energy transfer at 1.0~meV. \label{E_1p000_v}}
  \end{subfigure}
  
  \begin{subfigure}[b]{\textwidth}
    \includegraphics[width=90mm]{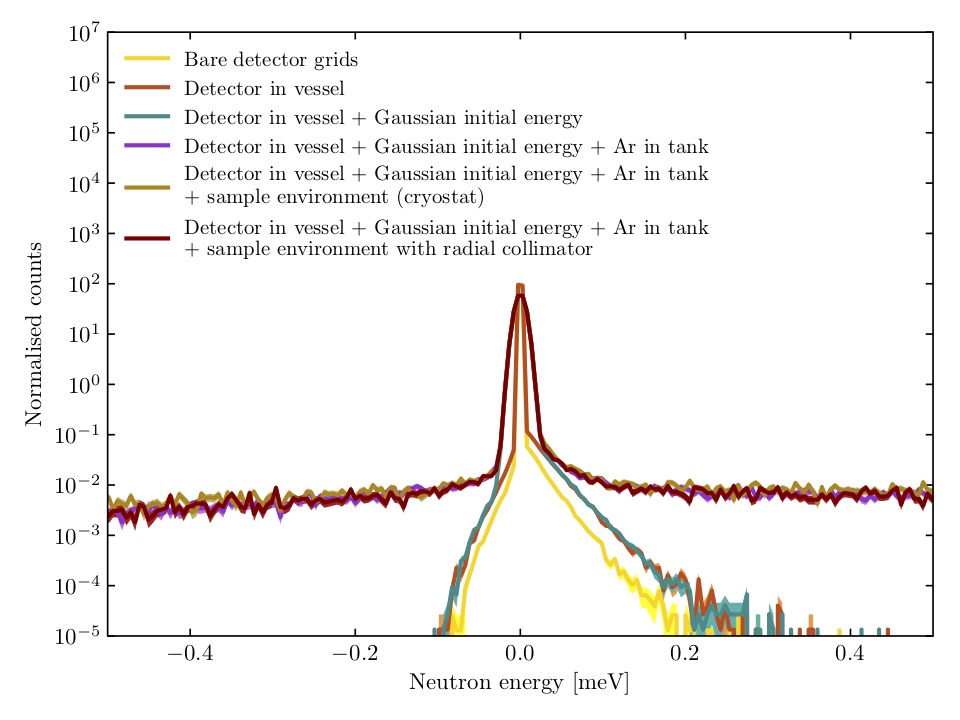}
    \caption{Energy transfer at 1.0~meV. \label{E_1p000_bg}}
  \end{subfigure}
  
  \caption{Measured and simulated energy transfer at 1.0~meV incident neutron energy~(\ref{E_1p000_v}) and comparison of the effect of different geometrical and instrumental parameters on energy transfer~(\ref{E_1p000_bg}). Energy transfer spectra are normalised to area.  \label{1p000_E_trf}}
  %%Data for comparison are measured on CNCS at SNS under ID IPTS-17219. \label{1p000_E_trf}}
\end{figure}
\begin{figure}[ht!]
  \centering
  
  \begin{subfigure}[b]{\textwidth}
    \includegraphics[width=90mm]{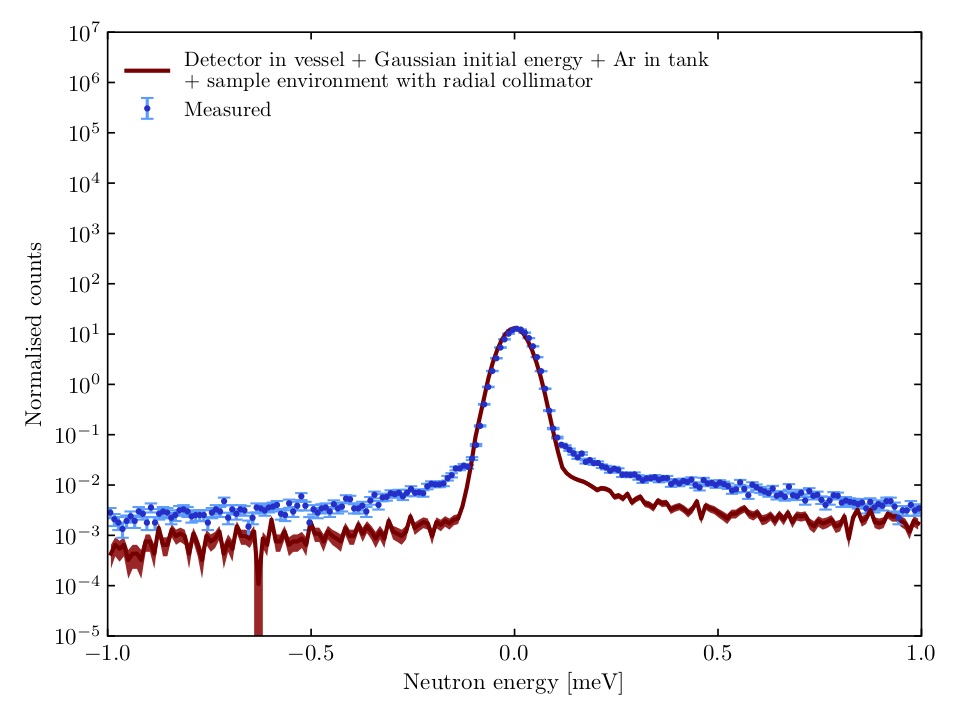}
    \caption{Energy transfer at 3.678~meV. \label{E_3p678_v}}
  \end{subfigure}
  
  \begin{subfigure}[b]{\textwidth}
    \includegraphics[width=90mm]{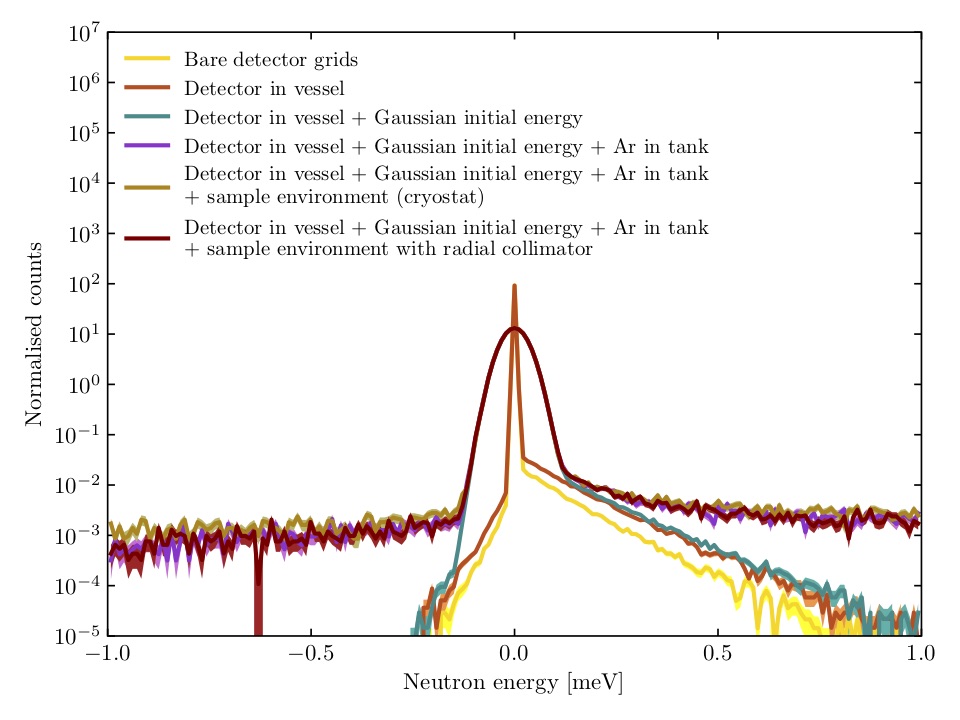}
    \caption{Energy transfer at 3.678~meV. \label{E_3p678_bg}}
  \end{subfigure}
  
  \caption{Measured and simulated energy transfer at 3.678~meV incident neutron energy~(\ref{E_3p678_v}) and comparison of the effect of different geometrical and instrumental parameters on energy transfer~(\ref{E_3p678_bg}). Energy transfer spectra are normalised to area. \label{3p678_E_trf}}
  %%Data for comparison are measured on CNCS at SNS under ID IPTS-17219. \label{3p678_E_trf}}
\end{figure}
%
%\FloatBarrier
%
%% \begin{figure}[ht!]
%%   \centering
%%
%%   \begin{subfigure}[b]{0.85\textwidth}
%%     \includegraphics[width=\textwidth]{mg_plots/3p807_E_trf_nat_sim_m.eps}
%%     \caption{Energy transfer at 3.807~meV. \label{E_3p807_v}}
%%   \end{subfigure}
%%  
%%   \begin{subfigure}[b]{0.85\textwidth}
%%     \includegraphics[width=\textwidth]{mg_plots/3p807_E_trf_nat_sim.eps}
%%     \caption{Energy transfer at 3.807~meV. \label{E_3p807_bg}}
%%   \end{subfigure}
%%
\begin{figure}[ht!]
  \centering

  \begin{subfigure}[b]{\textwidth}
    \includegraphics[width=90mm]{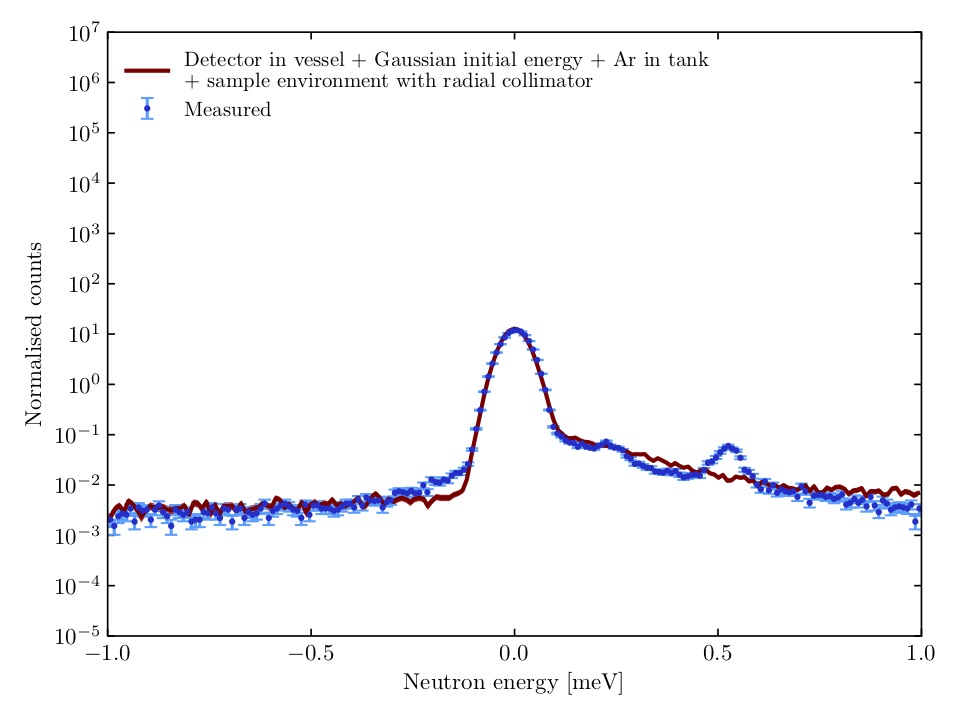}
    \caption{Energy transfer at 3.807~meV. \label{E_3p807_v}}
  \end{subfigure}
  
  \begin{subfigure}[b]{\textwidth}
    \includegraphics[width=90mm]{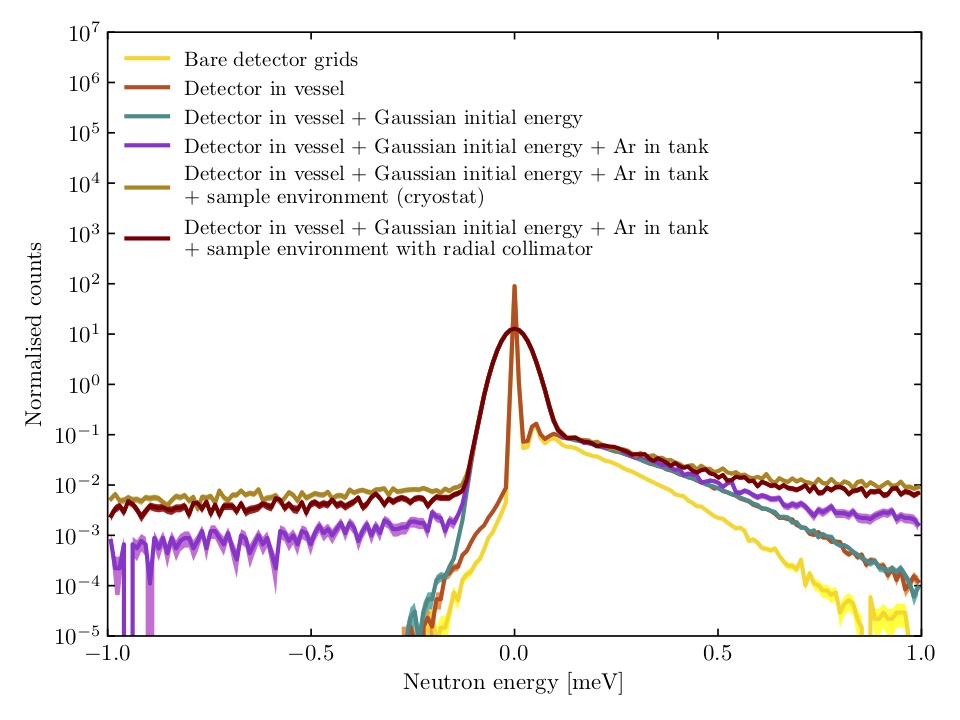}
    \caption{Energy transfer at 3.807~meV. \label{E_3p807_bg}}
  \end{subfigure}
  
  \caption{Measured and simulated energy transfer at 3.807~meV incident neutron energy~(\ref{E_3p807_v}) and comparison of the effect of different geometrical and instrumental parameters on energy transfer~(\ref{E_3p807_bg}). Energy transfer spectra are normalised to area. \label{3p807_E_trf}}
    %%Data for comparison are measured on CNCS at SNS under ID IPTS-17219. \label{3p807_E_trf}}
\end{figure}

\begin{sidewaystable}[htbp]
  \centering
  \caption{Simulated scattered neutron background ratio, normalised to
    elastic peak area (See Equation~\ref{eq:bg}). %Background iss defined as Total counts -- Counts in peak, where t
    The peak is defined in two ways: manually fitted peak width, and fix $\Delta$0.052~meV and $\Delta$0.24~meV peak width, equal to peak width of results with  $\rm E_{ini,Gaussian}$ for 1~meV and 3.678 and 3.807~meV, respectively.}
  \label{tab:bg_E_trf}
  \resizebox{\textwidth}{!}{
    \begin{tabular}{lcccccc}
      \hline
                        &  \multicolumn{2}{c}{1.0~meV}                     &  \multicolumn{2}{c}{3.678~meV}                      &  \multicolumn{2}{c}{3.807~meV}                       \\
      Model components  &  Adaptive peak width  & Fix $\pm$~0.026~meV width   &  Adaptive peak width  & Fix $\pm$~0.12~meV width    &  Adaptive peak width  & Fix $\pm$~0.12~meV width     \\
                        & Background ratio [\%] & Background ratio [\%]      & Background ratio [\%] & Background ratio [\%]       & Background ratio [\%] & Background ratio [\%]      \\
      \hline
      %% Bare grids & 0.23 & 0.06 & 1.79 & 0.82 \\
      %% Detector in vessel & 0.40 & 0.13 & 2.79 & 1.59 \\
      %% Detector in vessel + Gaussian $\rm E_{ini}$ & 0.15 & 0.15 & 1.56 & 1.56 \\
      %% Detector in vessel + Gaussian $\rm E_{ini}$ + Ar in tank & 1.43 & 1.42 & 2.58 & 2.55 \\
      %% Detector in vessel + Gaussian $\rm E_{ini}$ + Ar in tank + sample environment & 1.43 & 1.04 & 6.61 & 6.58
      %%REVISION
      %% Bare grids & 0.18 & 0.04 & 0.23 & 0.06 & 1.79 & 0.82 \\
      %% Detector in vessel & 0.39 & 0.12 & 0.40 & 0.13 & 2.79 & 1.59 \\
      %% Detector in vessel + Gaussian $\rm E_{ini}$ & 0.12 & 0.12 & 0.15 & 0.15 & 1.56 & 1.56 \\
      %% Detector in vessel + Gaussian $\rm E_{ini}$ & 1.33 & 1.33 & 1.43 & 1.42 & 2.58 & 2.55 \\
      %% + Ar in tank                                & &  &      &      &      &      \\ 
      %% Detector in vessel + Gaussian $\rm E_{ini}$ & 1.73 & 1.72& 1.43 & 1.04 & 6.61 & 6.58 \\
      %% + Ar in tank + sample environment           & & &      &      &      &
      Bare grids                                         & 0.18 & 0.04   & 0.23 & 0.06   & 1.80 & 0.82 \\
      Detector in vessel                                 & 0.39 & 0.12   & 0.42 & 0.13   & 2.81 & 1.60 \\
      Detector in vessel + Gaussian $\rm E_{ini}$         & 0.14 & 0.12   & 0.16 & 0.16   & 1.70 & 1.57 \\
      Detector in vessel + Gaussian $\rm E_{ini}$         & 1.31 & 1.32   & 1.10 & 1.09   & 2.64 & 2.57 \\
      + Ar in tank                                       & &  &      &      &      &      \\ 
      Detector in vessel + Gaussian $\rm E_{ini}$         & 1.72 & 1.71   & 1.42 & 1.42   & 6.90 & 6.62 \\
      + Ar in tank + sample environment                  & & &      &      &      &       \\
      Detector in vessel + Gaussian $\rm E_{ini}$        & 1.30 & 1.26   & 1.08 & 1.07   & 5.11 & 4.88 \\
      + Ar in tank + sample environment with collimator   & & &      &      &      & 
  \end{tabular}}

  \bigskip \bigskip  \bigskip

  \caption{Simulated scattered neutron background ratio, normalised to elastic peak area (See Equation~\ref{eq:bg}). %Background was defined as Total counts -- Counts in peak, where
    The peak was defined with manually fitted peak width.}
  \label{tab:bg_win}
  \resizebox{\textwidth}{!}{
    \begin{tabular}{cccccccc}
      \hline
       & & &                                                 & 1.0~meV               & 3.678~meV             & 3.807~meV             & 8.0~meV                       \\
      Vessel window & Grid entry & Vessel side & Vessel end  & Background ratio [\%] & Background ratio [\%] & Background ratio [\%] & Background ratio [\%]      \\
      %% \hline
      %%   0 mm  &  0.5 mm & 0 mm  &  0 mm & 0.18 & 0.22 & 1.77 & 2.07 \\ 
      %%   0 mm  &  0.5 mm & 3 mm  &  0 mm & 0.26 & 0.32 & 2.67 & 3.12 \\
      %%   0 mm  &  2 mm   & 3 mm  &  0 mm & 0.29 & 0.35 & 2.67 & 3.38 \\
      %%   3 mm  &  2 mm   & 3 mm  &  0 mm & 0.37 & 0.40 & 2.79 & 3.86 \\
      %%  20 mm  &  2 mm   & 3 mm  &  0 mm & 0.66 & 0.66 & 3.54 & 5.98 \\
      %%  3 mm  &  2 mm   & 3 mm  & 10 mm & 0.35 & 0.41 & 2.79 & 3.86
      %%REVISION
      \hline
       0 mm  &  0.5 mm & 0 mm  &  0 mm & 0.18   & 0.23   & 1.80   & 2.32 \\ 
       0 mm  &  0.5 mm & 3 mm  &  0 mm & 0.30   & 0.33   & 2.63   & 3.20 \\
       0 mm  &  2 mm   & 3 mm  &  0 mm & 0.33   & 0.36   & 2.70   & 3.47 \\
       3 mm  &  2 mm   & 3 mm  &  0 mm & 0.39   & 0.41   & 2.79   & 3.84 \\
      20 mm  &  2 mm   & 3 mm  &  0 mm & 0.73   & 0.68   & 3.58   & 5.91 \\
       3 mm  &  2 mm   & 3 mm  & 10 mm & 0.39   & 0.41   & 2.81   & 3.86 
  \end{tabular}}
\end{sidewaystable}

\subsection{Optimisation of the detector vessel window}\label{cncswin}

%DE
%% (KK: change first sentence or remove it)
%% The utilisation of the built
%% and now validated Geant4 Multi-Grid detector model was already
%% started.

A study is performed with the CNCS demonstrator model on the
effect of the aluminium window thickness and the vessel
components. The window thickness is defined as the sum of the vessel
window and the entry grid thickness. The 0.5~mm grid entry thickness
relates to the B$_{4}$C-coated blade, while bigger thicknesses
indicate the presence of an additional entry blade.  The effects of
the other parts of the vessel, the side and the rear end are also
considered. These components either appear with their realistic
dimensions or are removed. Combination of thicknesses are tested and
compared in the energy range of 1.0--8.0~meV in Figure~\ref{win}. The
set of simulated setups and the obtained backgrounds are presented in
Table~\ref{tab:bg_win}. The simulations are performed with
mono-energetic incident neutrons irradiating the entire detector
volume. Sample environment and tank gas are not present. 
\FloatBarrier
%
%% \begin{sidewaysfigure}[ht!]
%%   \centering
%%   \begin{subfigure}[b]{0.49\textwidth}
%%     \includegraphics[width=\textwidth]{mg_plots/1p000_win.eps}
%%     \caption{Energy transfer at 1.0~meV. \label{1p000_win}}
%%   \end{subfigure}
%%   %%
%%   \begin{subfigure}[b]{0.49\textwidth}
%%     \includegraphics[width=\textwidth]{mg_plots/3p678_win.eps}
%%     \caption{Energy transfer at 3.678~meV. \label{3p678_win}}
%%   \end{subfigure}
%%  
  %% \begin{subfigure}[b]{0.49\textwidth}
  %%   \includegraphics[width=\textwidth]{mg_plots/3p807_win.eps}
  %%   \caption{Energy transfer at 3.807~meV. \label{3p807_win}}
  %% \end{subfigure}
  %% %%
  %%   \begin{subfigure}[b]{0.49\textwidth}
  %%   \includegraphics[width=\textwidth]{mg_plots/8p000_win.eps}
  %%   \caption{Energy transfer at 8.0~meV. \label{8p000_win}}
  %% \end{subfigure}
\begin{sidewaysfigure}[ht!]
  \centering
  \begin{subfigure}[b]{0.49\textwidth}
    \includegraphics[width=90mm]{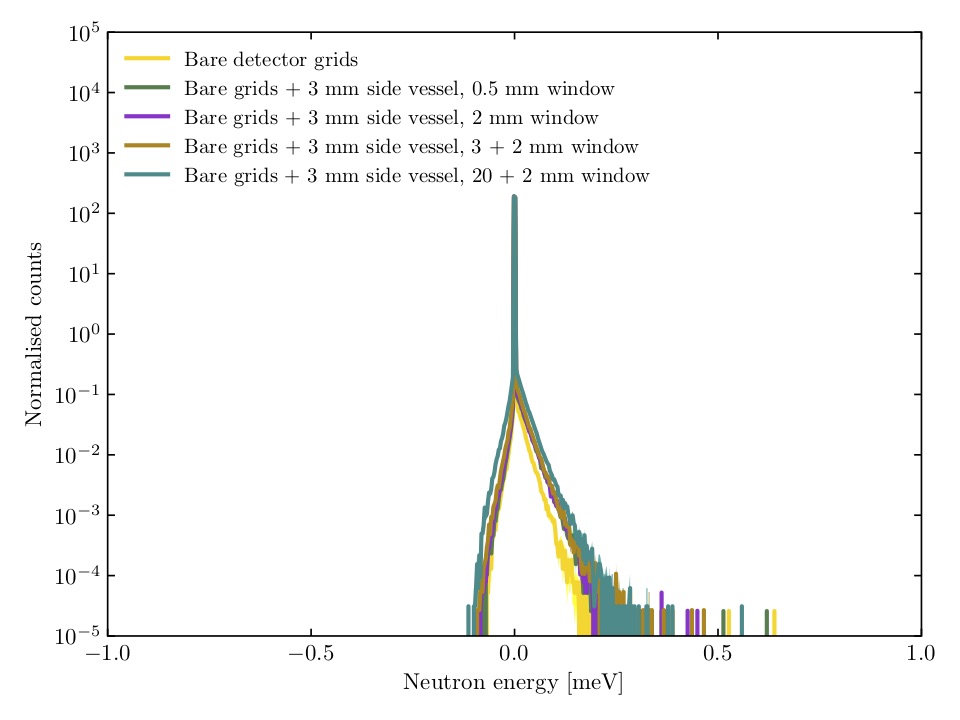}
    \caption{Energy transfer at 1.0~meV. \label{1p000_win}}
  \end{subfigure}
  \begin{subfigure}[b]{0.49\textwidth}
    \includegraphics[width=90mm]{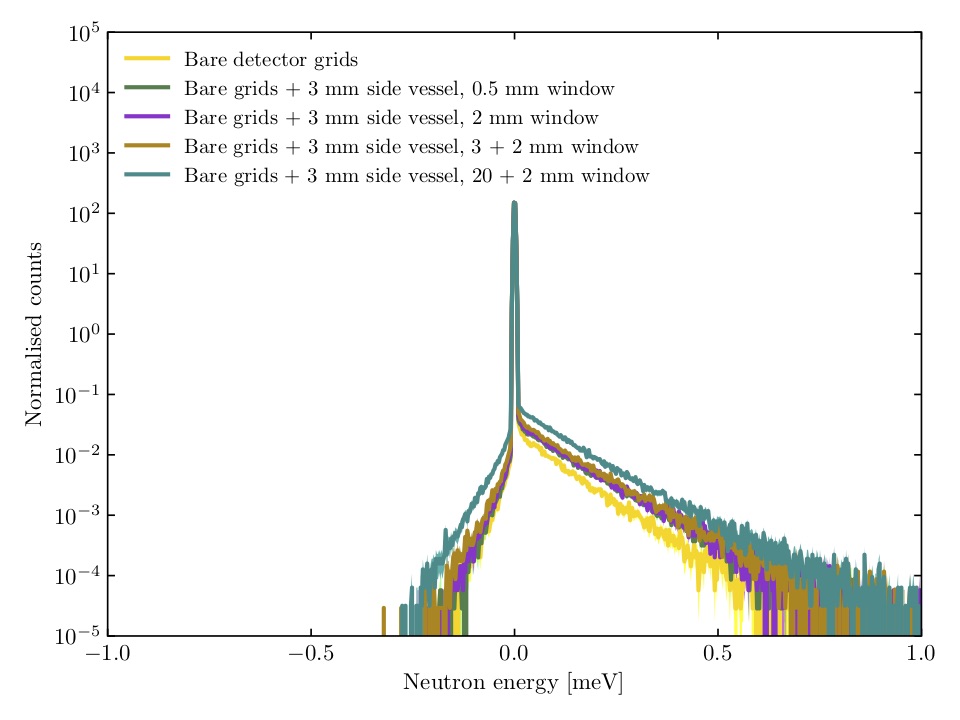}
    \caption{Energy transfer at 3.678~meV. \label{3p678_win}}
  \end{subfigure}
  
  \begin{subfigure}[b]{0.49\textwidth}
    \includegraphics[width=90mm]{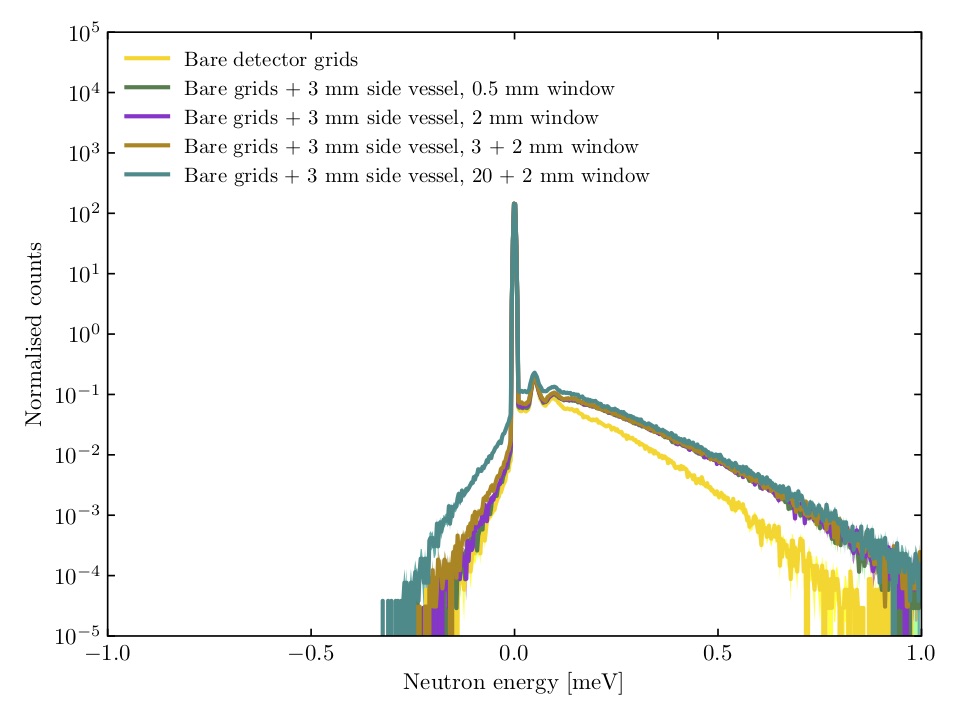}
    \caption{Energy transfer at 3.807~meV. \label{3p807_win}}
  \end{subfigure}
    \begin{subfigure}[b]{0.49\textwidth}
    \includegraphics[width=90mm]{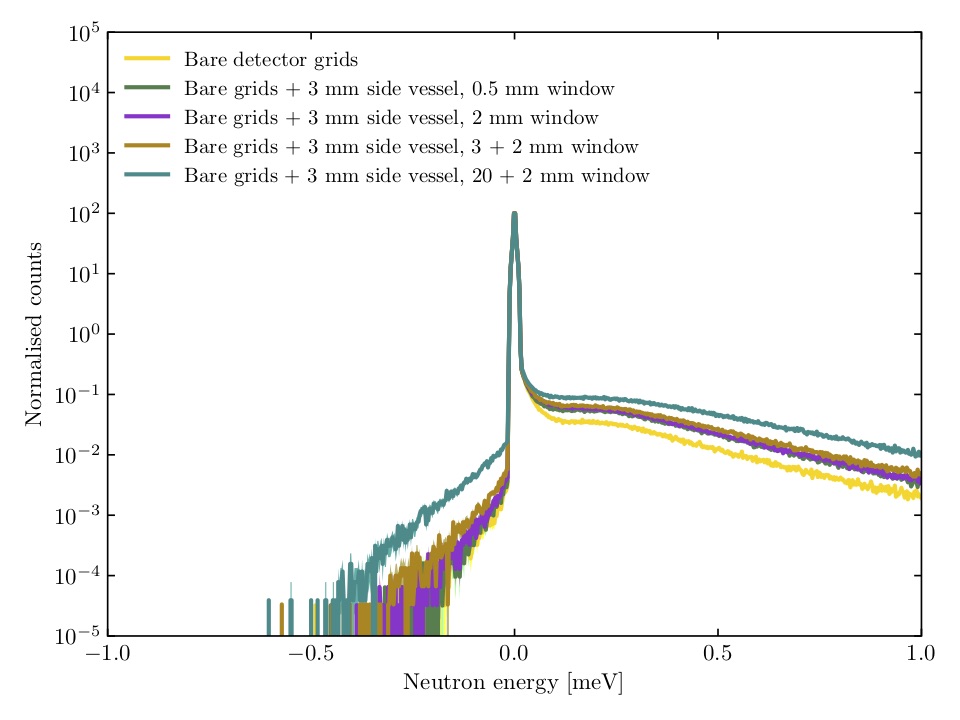}
    \caption{Energy transfer at 8.0~meV. \label{8p000_win}}
  \end{subfigure}
 
  \caption{Simulated energy transfer.  with the presence of different  window thicknesses and vessel components. Results are normalised to area. \label{win}}
\end{sidewaysfigure}

Comparing the results in the whole energy range it is shown that
except for the 22~mm total window thickness, which is %%an unrealistic extremity,
unrealistically thick, the difference in the background is negligible. However,
the presence of the side wall causes a significant increase in the
background on the positive side of the spectrum. Therefore, a
realistically chosen window thickness practically does not change the
scattered neutron background, but the application of shielding on
the inner wall of the vessel  might be considered.
%DE
%% the proper choice and the shielding
%% of the vessel side wall might be considered. (KK: do you mean material
%% choice?, composition and placement attributes?)

%% \paragraph{Experimentation}

%% \begin {enumerate}
%% \item CNCS
%% \item IN6
%% \end {enumerate}

%% \paragraph{FoMs? here or in Resultd \& Discussion?}

%%\section{Results \& Discussion}  

\section{Conclusions}\label{conc}

This is the first time sources of thermal neutron scattering background are modelled in a detailed simulation of detector response.

A detailed, realistic and flexible Geant4 model of the Multi-Grid
detector is built within the ESS Detector Group Simulation
Framework. The model is validated against measured data from the
demonstrators tested at the IN6 at ILL and the CNCS at SNS. 
%% The precise six-column
%% IN6 and the two-column CNCS demonstrator geometries are derived from
%% the general model. (KK: last sentence not necessary)
Measured ToF data are reproduced for the IN6 experiment both qualitatively (ToF - detection depth spectra) and quantitatively (ToF spectra).
The validated model is adopted for %%the more complex CNCS demonstrator detector, where instrument components are also included in the built geometry.
a more extensive set of measurements using a Multi-Grid detector at CNCS, including a more complete setup description.
The model is verified with the comparison of measured and simulated ToF and flight distance data at 3.678 and 3.807~meV (below and above the aluminium Bragg-edge).

A study is performed with the CNCS model to distinguish the sources of scattered neutron background. The %%REVIESD inelastic
elastic peak and the scattered neutron background in the energy transfer are now well-described and well-reproduced with the model, implying the predictive power of the simulation.

The simulation reveals that the neutron scattering in the detector geometry is minor in comparison with the effect of the scattering on instrument components: the tank gas and the sample environment; these are the major sources of the measured continuous flat background. 
%%REVISED
The sample environment should also be considered in the recently built instruments, operating with vacuum tank.
The effect of the detector window thickness is also studied in the range of 0.5 -- 22~mm. %% thicknesses.
It is shown that there is no significant change in the scattered neutron background for reasonable window thicknesses. The side of the vessel turns out to be a higher source of scattered neutron background, that should be taken into account in the further designs.

The availability of such a simulation allows to build neutron scattering instruments with optimised Signal-to-Background ratio by design.

\section*{Acknowledgements}

This work has been supported by the In-Kind collaboration between ESS~ERIC and the Hungarian Academy of Sciences, Centre for Energy Research (MTA~EK). Richard Hall-Wilton, Anton Khaplanov and Thomas Kittelmann would like to acknowledge support from the EU Horizon2020 Brightness Grant~[grant number 676548]. The authors would like to acknowledge the ILL and the SNS for the measured data. 
A portion of this research used resources at the Spallation Neutron Source, a DOE Office of Science User Facility operated by the Oak Ridge National Laboratory. 
CNCS data was measured at SNS under ID IPTS-17219. Computing resources provided by DMSC Computing Centre (https://europeanspallationsource.se/data-management-software/computing-centre).

%%\section*{References}

\newpage

%%\bibliography{../Reference}
%%\bibliography{Reference}

\begin{thebibliography}{10}
\expandafter\ifx\csname url\endcsname\relax
  \def\url#1{\texttt{#1}}\fi
\expandafter\ifx\csname urlprefix\endcsname\relax\def\urlprefix{URL }\fi
\expandafter\ifx\csname href\endcsname\relax
  \def\href#1#2{#2} \def\path#1{#1}\fi

\bibitem{ExpNSc2009}
B.~T.~M. Willis, C.~J. Carlile, {N}eutron {S}pectroscopy, in: B.~T.~M. Willis,
  C.~J. Carlile (Eds.), {E}xperimental {N}eutron {S}cattering, Oxford
  University Press, Oxford, 2009, Ch. 14-16, pp. 249--309.

\bibitem{mutka}
J.~Ollivier, H.~Mutka, L.~Didier, The new cold neutron time-of-flight
  spectrometer in5, Neutron News 21~(2) (2010) 22--25.

\bibitem{granroth2006}
G.~E. Granroth, et~al., {SEQUOIA}: {A} fine resolution chopper spectrometer at
  the {SNS}, {P}hysica {B}: {C}ondensed {M}atter 385-386, Part 2 (2006) pp.
  1104--1106.
\newblock \href {http://dx.doi.org/10.1016/j.physb.2006.05.379}
  {\path{doi:10.1016/j.physb.2006.05.379}}.

\bibitem{ehlers2011}
G.~Ehlers, et~al., {T}he {N}ew {C}old {N}eutron {C}hopper {S}pectrometer at the
  {S}pallation {N}eutron {S}ource: {D}esign and {P}erformance, {R}eview of
  {S}cientific {I}nstruments 82~(085108).
\newblock \href {http://dx.doi.org/10.1063/1.3626935}
  {\path{doi:10.1063/1.3626935}}.

\bibitem{ehlers2016}
G.~Ehlers, et~al., {T}he {N}ew {C}old {N}eutron {C}hopper {S}pectrometer at the
  {S}pallation {N}eutron {S}ource - a review of the first 8 years of operation,
  {R}eview of {S}cientific {I}nstruments 87~(093902).
\newblock \href {http://dx.doi.org/10.1063/1.4962024}
  {\path{doi:10.1063/1.4962024}}.

\bibitem{kajimoto2011}
R.~Kajimoto, et~al., {T}he {F}ermi {C}hopper {S}pectrometer 4{SEASONS} at
  {J-PARC}, {J}. {P}hys. {S}oc. {J}pn. 80~(SB025).
\newblock \href {http://dx.doi.org/10.1143/JPSJS.80SB.SB025}
  {\path{doi:10.1143/JPSJS.80SB.SB025}}.

\bibitem{nakajima2011}
K.~Nakajima, et~al., {AMATERAS}: {A} {C}old-{N}eutron {D}isk {C}hopper
  {S}pectrometer, {J}. {P}hys. {S}oc. {J}pn. 80~(SB028).
\newblock \href {http://dx.doi.org/10.1143/JPSJS.80SB.SB028}
  {\path{doi:10.1143/JPSJS.80SB.SB028}}.

\bibitem{bewley2011}
R.~I. Bewley, et~al., {LET}, a {C}old {N}eutron {M}ulti-{D}isk {C}hopper
  {S}pectrometer at {ISIS}, {N}uclear {I}nstruments and {M}ethods in {P}hysics
  {R}esearch Vol. 637, Issue 1 (2011) pp. 128--134.
\newblock \href {http://dx.doi.org/10.1016/j.nima.2011.01.173}
  {\path{doi:10.1016/j.nima.2011.01.173}}.

\bibitem{shea}
D.~Shea, D.~Morgan, {T}he {H}elium-3 shortage: supply, demand, and options for
  congress, Tech. Rep. R41419, {C}ongressional {R}esearch {S}ervice (December
  2010).

\bibitem{zeitelhack2012}
K.~Zeitelhack, {ICND}, Neutron News 23~(4) (2012) 10--13.

\bibitem{andersen_2012}
K.~Andersen, et~al., 10{B} multi-grid proportional gas counters for large area
  thermal neutron detectors, Nucl. Instr. and Meth. A 720 (2013) 116--121.

\bibitem{A.Khaplanov2012}
A.Khaplanov, 10b multi-grid proportional gas counters for large area thermal
  neutron detectors, Neutron News 23 (2012) 25.

\bibitem{ess}
\href{http://europeanspallationsource.se/}{{E}uropean {S}pallation {S}ource
  {ESS} {ERIC}}.
\newline\urlprefix\url{http://europeanspallationsource.se/}

\bibitem{deen2015}
P.~P. Deen, et~al., {A} design study of {VOR}: a {V}ersatile {O}ptimal
  {R}esolution chopper spectrometer for the {ESS}, {EPJ} {W}eb of {C}onferences
  83~(03002).
\newblock \href {http://dx.doi.org/10.1051/epjconf/20158303002}
  {\path{doi:10.1051/epjconf/20158303002}}.

\bibitem{TREXprop}
T.~Br\"uckel, J.~Voigt, N.~Violini, A.~Orecchini, A.~Paciaroni, F.~Sacchetti,
  M.~Zanatt, {ESS} {I}nstrument {C}onstruction {P}roposal {T-REX}: {A}
  {T}ime-of-flight {R}eciprocal space {E}xplorer.

\bibitem{CSPECprop}
W.~Lohstroh, W.~Petry, J.~Neuhaus, L.~Silvi, C.~Alba-Simionesco, J.-M. Zanotti,
  S.~Longeville, {ESS} {I}nstrument {C}onstruction {P}roposal {C-SPEC} - {C}old
  chopper spectrometer.

\bibitem{mg_patent_ill}
B.~Guerarad, J.-C. Buffet, no. US~2011215251, (Laue Max Inst, France), 2010.

\bibitem{ill}
\href{http://www.ill.eu}{{I}nstitute {L}aue-{L}angevin}.
\newline\urlprefix\url{http://www.ill.eu}

\bibitem{crisp}
\href{www.crisp-fp7.eu}{{T}he {C}luster of {R}esearch {I}nfrastructures for
  {S}ynergies in {P}hysics}.
\newline\urlprefix\url{www.crisp-fp7.eu}

\bibitem{brightness}
\href{https://brightness.esss.se/}{{B}rightn{ESS}}.
\newline\urlprefix\url{https://brightness.esss.se/}

\bibitem{hoglund2012}
C.~H\"{o}glund, {B4C} thin films for neutron detection, J. Appl. Phys. 111
  (2012)~(104908).

\bibitem{c.hoglund2015b}
C.~H\"{o}glund, et~al., Stability of $^{10}${B}$_{4}${C} thin films under
  neutron radiation, Radiation Physics and Chemistry 113 (2015) 14--19.

\bibitem{LiU2016b}
S.~Schmidt, et~al., Low-temperature growth of boron carbide coatings by direct
  current magnetron sputtering and high-power impulse magnetron sputtering,
  Journal of Materials Science 51.
\newblock \href {http://dx.doi.org/10.1007/s10853-016-0262-4}
  {\path{doi:10.1007/s10853-016-0262-4}}.

\bibitem{khaplanov2014}
A.~Khaplanov, et~al., In-beam test of the {B}oron-10 {M}ulti-{G}rid neutron
  detector at the {IN6} time-of-flight spectrometer at the {ILL}, J. Phys.:
  Conf. Ser. 528 (2014)~(012040).

\bibitem{khaplanov2017}
A.~Khaplanov, et~al., {M}ulti-{G}rid {D}etector for {N}eutron {S}pectroscopy:
  {R}esults {O}btained on {T}ime-of-{F}light {S}pectrometer {CNCS}, JINST 12
  (2017) P04030.

\bibitem{agostinelli}
S.~Agostinelli, et~al., Geant4: A simulation toolkit, Nucl. Instr. and Meth. A
  506 (2003) 250--303.

\bibitem{allison2006}
J.~Allison, et~al., Geant4 developments and applications, {IEEE}
  {T}rans.{N}ucl. {S}ci. vol. 53~(no. 1) (2006) pp. 270--278.
\newblock \href {http://dx.doi.org/10.1109/TNS.2006.869826}
  {\path{doi:10.1109/TNS.2006.869826}}.

\bibitem{allison2016}
J.~Allison, et~al., Recent developments in geant4, {N}ucl. {I}nstrum. {M}ethods
  {P}hys. {R}es. {S}ection {A} {A}ccel. {S}pectrometers {D}etect. {A}ssoc.
  {E}quip. vol. 835 (2016) pp. 186--225.
\newblock \href {http://dx.doi.org/10.1016/j.nima.2016.06.125}
  {\path{doi:10.1016/j.nima.2016.06.125}}.

\bibitem{kittelmann2013b}
T.~Kittelmann., et~al., Geant4 based simulations for novel neutron detector
  development, {J}. {P}hys.: {C}onf. {S}er Vol. 513.
\newblock \href {http://dx.doi.org/doi:10.1088/1742-6596/513/2/022017}
  {\path{doi:doi:10.1088/1742-6596/513/2/022017}}.

\bibitem{kittelmann2015b}
T.~Kittelmann, M.~Boin, Polycrystalline neutron scattering for {Geant4}:
  {NXSG4}, {C}omput. {P}hys. {C}ommun. 189 (2015) pp. 114--118.
\newblock \href {http://dx.doi.org/doi:10.1016/j.cpc.2014.11.009}
  {\path{doi:doi:10.1016/j.cpc.2014.11.009}}.

\bibitem{kittelmann2017}
T.~Kittelmann, K.~Kanaki, E.~Klinkby, X.~X. Cai, C.~P. Cooper-Jensen,
  R.~Hall-Wilton, Using backscattering to enhance efficiency in neutron
  detectors, {IEEE} {T}ransactions on {N}uclear {S}cience 64~(6) (2017) pp.
  1562--1573.
\newblock \href {http://dx.doi.org/10.1109/TNS.2017.2695404}
  {\path{doi:10.1109/TNS.2017.2695404}}.

\bibitem{ref4Bragg}
S.~T. Thornton, A.~Rex, Modern Physics for Scientists and Engineer, Saunders
  College Publishing, 1993.

\bibitem{AlCrystal}
E.~A. Owen, E.~L. Yates, Precision measurements of crystal parameters,
  {P}hilosophical {M}agazine 15 (1933) pp. 472--488.

\bibitem{birch2015a}
J.~Birch, et~al., Investigation of background in large-area neutron detectors
  due to alpha emission from impurities in aluminium, Journal of
  Instrumentation 10 (2015) P10019.
\newblock \href {http://dx.doi.org/10.1088/1748-0221/10/10/P10019}
  {\path{doi:10.1088/1748-0221/10/10/P10019}}.

\bibitem{peters2017}
J.~Peters, J.~D.~M. Champion, G.~Zsigmond, H.~N. Bordallo, F.~Mezei, Using
  {F}ermi choppers to shape the neutron pulse, {NIM} {A} 557~(2) (2006) pp.
  580--584.
\newblock \href {http://dx.doi.org/10.1016/j.nima.2005.11.111}
  {\path{doi:10.1016/j.nima.2005.11.111}}.

\bibitem{mirrobor}
\href{https://mirrotron.com/en/products/radiation-shielding}{{M}irrotron
  {R}adiation {S}hielding}.
\newline\urlprefix\url{https://mirrotron.com/en/products/radiation-shielding}

\bibitem{khaplanov2013a}
A.~Khaplanov, et~al., Investigation of gamma-ray sensitivity of neutron
  detectors based on thin converter films, JINST 8 (2013)~(P10025).
\newblock \href {http://dx.doi.org/10.1088/1748-0221/8/10/P10025}
  {\path{doi:10.1088/1748-0221/8/10/P10025}}.

\bibitem{mauri2018}
G.~Mauri, et~al., Fast neutron sensitivity of neutron detectors based on
  boron-10 converter layers, {J}ournal of {I}nstrumentation 13 (2018) P03004.
\newblock \href {http://dx.doi.org/10.1088/1748-0221/13/03/P03004}
  {\path{doi:10.1088/1748-0221/13/03/P03004}}.

\bibitem{catia6}
\href{https://www.3ds.com/products-services/catia/products/v6/}{{CATIA}6}.
\newline\urlprefix\url{https://www.3ds.com/products-services/catia/products/v6/}

\end{thebibliography}

\end{document}